\documentclass[preprint,sort&compress,number,eqsecnum]{elsarticle}
\def \bfgr #1{ \mbox {{\boldmath $#1$}}}
\usepackage{axodraw}
\usepackage{threeparttable}
\usepackage{url}
\usepackage{bm}
\usepackage[breaklinks=true,colorlinks=true,urlcolor=cyan]{hyperref}
\begin{document}
\setcounter{page}{0}
\title{The effective description of non-strange hadrons
 low-energy electro-weak transitions.}
\author{G.  G.  Bunatian \footnote{Corresponding author.\\Email address : \href{mailto:bunat@cv.jinr.ru}{bunat@cv.jinr.ru} (G. Bunatian)}
\\{\it Joint Institute for Nuclear Research,
 141980, Dubna, Russia}}
\date{\today}
\begin{abstract}
 Starting with the general principles of global and local
symmetries, the effective pion-nucleon lagrangian, essentially
non-linear in pion field, to describe the non-strange hadrons
low-energy electro-weak transitions is developed. We encounter no
divergence summarizing properly all the infinite power series in
pion field which occur in the course of treatment. Our consistent
approach proves to be relevant in considering P-parity violation in
pion-nucleon interactions.\\

\vspace{0.2cm}

{PACS 11.10.Lm Nonlinear or nonlocal theories and models --- 11.15.Tk Other nonperturbative techniques --- 12.15.Ji Application of electroweak models to specific processes --- 11.30.Er Charge conjugation, parity, time reversal, and other discrete symmetries \\

\vspace{0.3cm}

}
\end{abstract}
\maketitle
\setcounter{page}{1}
\setcounter{section}{0}
\section{Introduction}
\setcounter{equation}{0}
\label{sec:level1}
 Nowadays, in actual treating low-energy processes in hadronic
systems, in particular the hadron electro-weak transitions, there
sees no option but having recourse to the effective lagrangian to
describe hadron structure and strong interactions,
\begin{equation}
{\cal L}_h\equiv{\cal L}_h[\{\phi_i(x)\},
\{\partial_{\mu}\phi_i(x)\}],
\label{00}
\end{equation}
where the hadron fields $\phi_i$ satisfy the Euler-Lagrange
equations, and the (anti)commutation relations hold for the
Heisenberg field operators $\phi_i$ and the corresponding
canonically-conjugated momenta (see, for instance, refs.
\cite{d,eft,cpt}).

Plain purpose of the work presented is to study the relevant
construction of the effective hadron lagrangian and the respective
hadron currents to describe the
 low-energy electro-weak transitions of non-strange hadrons. Thus
far, the Standard Model ($SM$) (the Weinberg-Salam theory), which
is written in terms of quark, gluon, and lepton degrees of freedom,
is hardy applied to this problem since we still lack a reliable and
calculable way to describe hadrons in terms of their quark-gluon
degrees of freedom. Although $QCD$ is widely understood to be the
correct theory of strong interaction, it is difficult to apply
rigorously away from the high-energy perturbative area. Since the
issue of treating the electro-weak hadronic transitions on the
basis of ``first principles" is not yet resolved, a calculation
starting with the $SM$ involves inevitably simplifying which causes
the considerable ambiguities in findings. Our way along, we
sidestep plunging into all the intricate $QCD$ analyzes.

Our treatment is based on the nonlinear realization of the chiral
symmetry of the pion-nucleon lagrangian, with this symmetry
partially broken. In our consideration, the nucleon and pion are
principal particles, and all the other hadronic states, particles
and resonances, are interpreted to have the dynamical origin.
Although we consider the system of non-strange hadrons stable
against strong decay modes, nucleons and pions, the lagrangian
${\cal L}_h$ is in general certain to describe the variety of
electro-weak transitions of non-strange hadrons and hadron
resonances, made an expedient allowance for the nucleon-pion strong
interaction.

In Sec. II, we treat the essentially nonlinear pion-nucleon
lagrangian, with the pion wavefunction and mass renormalization
properly carried out. In  \hyperlink{appen}{Appendix A}, the dimensional regularization
(or continuous dimension) method is considered to treat properly the ultraviolet divergent quantities which occur in our calculations.
In Sec. III, the conserved and
partially-conserved hadron currents, radically nonlinear in pion
field, associated with that lagrangian are derived. These currents
serve to construct, on the local symmetry principle ground, the
lagrangian to describe the hadron electro-weak transitions caused
by interactions with the electromagnetic field and the intermediate
gauge $W^{\pm}-$ and $Z^0-$ boson fields, Sec. IV. Upon specifying
all the parameters involved in the treatment, this lagrangian is
relevant to study the electro-weak processes of significant
interest to day. In particular, it provides the way of treating
P-parity violation in pion-nucleon interactions, as discussed in
Sec. V. In the last Sec. VI, the relevant concluding points are set
forth.

The strong interaction is presumed to be adequately allowed for as
our treatment is essentially nonlinear in pion field. In our
calculation, we are dealing with the infinite power series in pion
field. Without restricting by a finite number of terms, we
summarize all this infinite series in due course, no divergence
 emerging thereby, which validates our approach.

\section{Pion-nucleon effective lagrangian}
\label{sec:level2}

In order to embark on the path to the effective nonlinear
pion-nucleon lagrangian approach, we pursue the method developed in
the widely known investigations \cite{jm,gu,vp}. The nucleon
lagrangian to start with,
\begin{equation}
{\cal L}_{\psi}(x)= \bar\psi(x)[i\gamma^{\mu}\partial_{\mu} -
M]\psi(x) \; ,
\label{o1n}
\end{equation}
is invariant under the global $SU(2)$ isospin rotation
\begin{equation}
\psi(x)\rightarrow\psi'(x)=\mbox{e}^{-\frac{i}{2}\bm{\tau\varepsilon}}
 \psi(x) , \label{1e}
\end{equation}
where $\{\tau_i\} , \, (i=1,2,3)$ are the Pauli matrices.
 Yet, as the nucleon mass $M$ is involved, it is apparently
non-invariant under the global chiral transformation
\begin{equation}
\psi(x)\rightarrow\psi'(x)=\mbox{e}^{\frac{i}{2}\gamma^5
\bm{a\tau}} \psi(x) \,. \label{1a}
\end{equation}
To balance out the chirality violation, the nucleon interaction
with the compensating pseudo scalar field $\bm{\varphi}(x)$ is
introduced by the replacement
\begin{eqnarray}
{\cal L}_{\psi}\rightarrow{\cal
L}_{\psi\varphi}=\bar\psi[i\gamma^{\mu}\partial_{\mu}- M{\large\bf
U}]\psi=\bar\psi i\gamma^{\mu}\partial_{\mu}\psi-
\bar\psi M\psi+{\cal L}_{\psi\varphi}^{int} \, , \label{2l}\\
{\cal L}^{int}_{\psi\varphi}=M\bar\psi(1-{\large\bf U})\psi
\, ,
\label{i2l}
\end{eqnarray}
with ${\large\bf U}$ being a matrix function of
$i\gamma^5 \bm{\tau\varphi}$. Provided the transformations
\begin{eqnarray}
{\large\bf U}\rightarrow{\large\bf U}'=\mbox{e}^{-\frac{i}{2}
\bm{\varepsilon\tau}}{\large\bf U}\mbox{e}^{\frac{i}{2}
\bm{\varepsilon\tau}} \, , \label{1u}\\
{\large\bf U}\rightarrow{\large\bf
U}'=\mbox{e}^{-\frac{i}{2}\gamma^5
\bm{a\tau}}{\large\bf U}\mbox{e}^{-\frac{i}{2}\gamma^5
\bm{a\tau}} \,  \label{2u}
\end{eqnarray} are constrained when the nucleon field transforms
by eqs. (\ref{1e}), (\ref{1a}), the lagrangian (\ref{2l}) proves to
be invariant under the global $SU(2)\times SU(2)$ isospin and
chiral transformation. As we purpose to study the low-energy hadron
electro-weak processes, the lagrangian of the pseudo-scalar
compensating field $\bm{\varphi}$ itself is written in the
simplest, two-derivative, invariant form
\begin{eqnarray}
{\cal L}_{\varphi\varphi}=\frac{1}{16f^2}\mbox{Sp}[\partial_{\mu}
{\large\bf U}\partial^{\mu}{\large\bf U}^{\dagger}] \, .
\label{1pi}
\end{eqnarray}
For non-linear realization of the isospin and chiral global
invariance of lagrangian, the specific form of the matrix
${\large\bf U}$ is chosen
\begin{equation}
{\large\bf U}=\mbox{e}^{-i2f\gamma^5\bm{\tau\varphi}} \, ,
\label{3u}
\end{equation}
so that ${\cal L}_{\varphi\varphi}$ (\ref{1pi}) will reduce to the
ordinary pseudo-scalar field kinetic energy
\begin{equation}
{\cal L}_{\varphi\varphi}\approx
\frac{1}{2}\partial_{\mu}\bm{\varphi}\partial^{\mu}\bm{\varphi}
 \, , \label{0pe}
\end{equation}
when the dimensional parameter $f$ tends to zero. The interaction
${\cal L}^{int}_{\psi\varphi}$ (\ref{i2l}) reduces in this limit as
follows
\begin{equation}
{\cal L}^{int}_{\psi\varphi}\approx
2ifM\bar\psi\gamma^5\bm{\tau\varphi}\psi \, . \label{nli}
\end{equation}
To associate $\bm{\varphi}$ with the real pion field of the mass
$m$, the lagrangian ${\cal L}_{\varphi\varphi}$ (\ref{1pi}) is
known to be accomplished
 \cite{gu,vp} by adding the chiral symmetry-breaking term
\begin{eqnarray}
{\cal L}_{SB}=\frac{m^2}{16f^2}\mbox{Sp}[{\large\bf U}_{\xi}+
{\large\bf U}^{\dagger}_{\xi}-2]=
-\frac{m^2}{2f^2}\sin^2(f\sqrt{\bm{\varphi
}^2})\approx
-\frac{m^2\bm{\varphi}^2}{2} + \frac{m^2f^2\bm{\varphi}^4}{6}\, ,
\label{sb}\\
 {\large\bf U}_{\xi}=\xi \xi \, , \; \; \xi=
\mbox{e}^{-i f\bm{\tau\varphi}} \, ,\nonumber
\end{eqnarray} so that PCAC  will fulfill.

By the canonical transformation \cite{d,gu,vp}
\begin{equation}
N={\large\bf U}_b^{-b/2}\psi \, , \; \; \; {\large\bf
U}_b^{b/2}=\mbox{e}^{i\gamma^5f \bm{\tau\varphi} b} \label{1ct}
\end{equation}
of the nucleon field, the original lagrangian ${\cal
L}_{\psi\varphi}$ (\ref{2l}), that involves only the non-derivative
pion-nucleon couplings, is transformed to the general form
\begin{equation}
{\cal L}_{\psi\varphi}\rightarrow{\cal L}_{bN\varphi}=i\bar
N\gamma^{\mu}\partial_{\mu}N+i\bar N\gamma^{\mu} {\large\bf
U}_b^{-b/2}\partial_{\mu}{\large\bf U}_b^{b/2} N - \bar N
{\large\bf U}_b^{b/2} {\large\bf U}{\large\bf U}_b^{b/2} N M ,
 \label{dnp}
\end{equation}
with $b$ being some dimensionless parameter. Then, with introducing
the auxiliary quantities
\begin{eqnarray}
\xi^b=\mbox{e}^{ib f \bm{\tau\varphi}} \, , \label{xi}\\
  V_{\mu}=\frac{i}{2}\Bigl( \xi^{-b}\partial_{\mu}\xi^{b}
  +\xi^{b}\partial_{\mu}\xi^{-b} \bigl) \, , \label{v}\\
 A_{\mu}=\frac{i}{2}\Bigl( \xi^{-b}\partial_{\mu}\xi^{b}
  -\xi^{b}\partial_{\mu}\xi^{-b} \bigl) \, , \label{a}
\end{eqnarray}
 the total $SU(2)\times SU(2)$ invariant lagrangian is convenient
to be generally rewritten as follows
\begin{eqnarray}
{\cal L}_h ={\cal L}_N +{\cal L}^{int}_{1 \, N\varphi} +{\cal
L}^{int}_{0 \, N\varphi} +{\cal L}_{\varphi\varphi} \label{4l} \,
,\\
 {\cal L}_N =i\bar N\gamma^{\mu}\partial_{\mu}N \, , \label{l00}\\
{\cal L}^{int}_{1 \, N\varphi}=\bar N\gamma^{\mu}V_{\mu}N +g'\bar
N\gamma^{\mu}A_{\mu}\gamma^5 N =\nonumber\\
=\bar N\gamma^{\mu}\Biggl\{-\bm{\tau}
[\bm{\varphi}\times\partial_{\mu}\bm{\varphi}]
\frac{\sin^2(z)}{\bm{\varphi}^2} -
g'\gamma^5\Bigl\{\frac{\bm{\tau}\partial_{\mu}\bm{\varphi}}
{2\sqrt{\bm{\varphi}^2}}\sin(2z)+\nonumber\\  +
 \frac{(\bm{\tau\varphi})
 (\bm{\varphi}\partial_{\mu}\bm{\varphi})} {\bm{\varphi}^2} f
 b \Bigl(1-\frac{\sin(2z)}{2z}\Bigr)\Bigr\}\Biggr\}N\approx
\label{il1}\\
\approx-g' (f b)  \bar N\gamma^{\mu}\gamma^5\bm{\tau}\partial_{\mu}
\bm{\varphi} N -f^2 b^2 \bar N \gamma^{\mu}
[\bm{\varphi}\times\partial_{\mu}\bm{\varphi}]\bm{\tau} N +
{\cal O}(f^3) \, , \; \; \; \; z=f b \sqrt{\bm{\varphi}^2} \, ,
\nonumber\\
 {\cal L}^{int}_{0 \, N\varphi}=
-\bar N [\exp(2i\gamma^5 f\bm{\tau\varphi}(b-1))]
N M\approx \nonumber\\
\approx-2if(b-1) M \bar
 N\gamma^5\bm{\tau\varphi} N +2f^2(b-1)^2\bm{\varphi}^2 \bar N
N M- \bar N N M + {\cal O}(f^2 ) \, , \label{il0} \\ {\cal
L}_{\varphi\varphi}=
\frac{1}{2}\partial_{\mu}\bm{\varphi}\partial^{\mu}
\bm{\varphi}\Bigl(\frac{\sin^2(y)}{y^2}\Bigr)+
\frac{(\bm{\varphi}\partial_{\mu}\bm{\varphi})^2}
{2\bm{\varphi}^2}
\Bigl(1-\frac{\sin^2(y)}{y^2}\Bigr)\approx\label{2pi}\\
 \approx\frac{1}{2}\partial_{\mu}\bm{\varphi}\partial^{\mu}
 \bm{\varphi} -
 \frac{2}{3}f^2\bm{\varphi}^2\Bigl(\partial_{\mu}\bm{\varphi}
 \partial^{\mu}\bm{\varphi}-\frac{(\bm{\varphi}\partial_{\mu}
 \bm{\varphi})^2}{\bm{\varphi}^2}\Bigr)+
  {\cal O}(f^4) \, , \; \; \; y=2f\sqrt{\bm{\varphi}^2} \, ,
\nonumber
\end{eqnarray}
 where introducing a parameter $g'$
  does not spoil invariance of the lagrangian \cite{d,gu,vp,kvp}.
With adding the term ${\cal L}_{SB}$ (\ref{sb}), the lagrangian
 ${\cal L}_{\varphi\varphi}$ (\ref{2pi}) transforms to
\begin{eqnarray}
{\cal L}_{\varphi\varphi \, SB}={\cal L}_{\varphi\varphi}+{\cal
L}_{SB}\approx{\cal L}_{\varphi\varphi}-
\frac{m^2}{2}\bm{\varphi}^2 + \frac{m^2}{6}f^2(\bm{\varphi
\varphi})^2
+{\cal O}(f^4) . \label{dd}
\end{eqnarray}

As seen, the lagrangian (\ref{4l}) incorporates both derivative and
non-derivative pion-nucleon couplings. The relation among these two
kinds of couplings is dictated by choice of $b$ value. Generally
speaking, the parameter $b$ could be considered as a fit-parameter
to be fixed by processing the appropriate experimental data. Chosen
hereafter $b=1$,
 we arrive at the most frequently considered lagrangian comprising
only the derivative pion-nucleon couplings, with ${\cal L}^{int}_{0
N \varphi}$ therein reduces merely to $-{\bar N}N M$. The effective
lagrangian (\ref{4l}) with $b=1$ is known to be obtained in the
general geometrical method \cite{kvp,dv,col}, and also deduced
starting with the $\sigma$-model when the $\sigma$-meson is finely
eliminated \cite{sig,d}. At $b=0$ we would apparently return to the
non-derivative couplings of pions and nucleons (\ref{2l}),
(\ref{i2l}), (\ref{nli}). In processing various experimental data,
the various suggestions are managed
 for purpose of best fitting. For instance, the succeeding
 canonical transformation of the nucleon field, with the parameter
 $g'$,
\begin{equation}
N_1\rightarrow\mbox{e}^{ig'f\gamma^5\bm{\tau\varphi}} N \,
\label{chn}
\end{equation}
was performed in the work \cite{eh} so as the effective
pion-nucleon lagrangian to work with would reduce to
\begin{eqnarray}
 {\cal L}_{N_1}+{\cal L}_{1
\, N_1\varphi}^{int} + {\cal L}_{0 \, N_1\varphi}^{int}
\approx -iMg'2f\bar
N_1\gamma^5\bm{\tau\varphi}N_1+Mg'^2 2f^2\bm{\varphi}^2\bar N_1
N_1 +\nonumber \\ +(g'^2-1)f^2\bar N_1
\gamma^{\mu}[\bm{\varphi}
\times\partial_{\mu}\bm{\varphi}]\bm{\varphi}
N_1 - \bar N_1 N_1 M + {\cal O}(f^3) \, . \label{dh}
\end{eqnarray}

Let us now treat the pion mass and wavefunction renormalization.
The lagrangian (\ref{dd}) is expedient to be presented as
\begin{eqnarray}
\bar{\cal L}_{\varphi\varphi SB} = {\cal L}_{\varphi\varphi}^0 +
\overline{\cal L}_{\varphi\varphi}^{int} ,\label{r0}\\
{\cal L}_{\varphi\varphi}^0
=\frac{1}{2}\partial_{\mu}\bm{\varphi}(x)\partial^{\mu}
\bm{\varphi}(x) - \frac{m^2}{2}\bm{\varphi}^2(x) ,\nonumber\\
\overline{\cal L}_{\varphi\varphi}^{int}=
{\cal M}\bm{\varphi}^2(x)+ {\cal
K}\partial_{\mu}\bm{\varphi}(x)\partial^{\mu}
\bm{\varphi}(x) ,\nonumber
\end{eqnarray}
with introducing the subsidiary quantities ${\cal K} ,
\, {\cal M}$. This lagrangian transforms immediately to the
canonical form as follows
\begin{eqnarray}
\bar{\cal L}_{\varphi\varphi SB}=
\frac{1}{2}\partial_{\mu}\bm{\varphi}(x)\partial^{\mu}
\bm{\varphi}(x)[1+2{\cal K}]-\frac{m^2}{2}\bm{\varphi}^2(x)
[1-\frac{2{\cal M}}{m^2}]= \nonumber\\
=\frac{1}{2}\partial_{\mu}\bm{\varphi}_r(x)\partial^{\mu}
\bm{\varphi}_r(x)-\frac{m^2}{2}\bm{\varphi}_r^2(x){\mathsf z}
[1-\frac{2{\cal M}}{m^2}]= \label{r2} \\
=\frac{1}{2}\partial_{\mu}\bm{\varphi}_r(x)\partial^{\mu}
\bm{\varphi}_r(x)-\frac{m_r^2}{2}\bm{\varphi}^2_r(x) ,
\nonumber
\end{eqnarray}
where the renormalized wavefunction $\bm{\varphi}_r$ and mass
$m_r$ are identified as
\begin{eqnarray}
\bm{\varphi}_r(x)=\bm{\varphi}(x){\mathsf z}^{-1/2} , \; \; {\mathsf
z}^{-1}=1+2{\cal K} , \; \; m_r^2=m^2{\mathsf z}\Bigl(1-\frac{2{\cal
M}}{m^2}\Bigr)=m^2\Bigl(1-2{\cal K}-\frac{2{\cal M}}{m^2}\Bigr). \; \; \; \; \; \; \; \;
\label{r3}
\end{eqnarray}
By correlating the propagators
\begin{eqnarray}
\langle 0|{\cal
T}{\varphi_{\alpha}(x_1)\varphi_{\beta}(x_2)}\exp[i\int\mbox {d}x^4
\hat{\cal L}_{\varphi\varphi}(x)]|0\rangle , \label{pr}
\end{eqnarray}
calculated with the interactions $\hat{\cal
L}_{\varphi\varphi}={\cal L}_{\varphi\varphi SB}-{\cal
L}^0_{\varphi\varphi}$ (\ref{dd}) and $ \hat{\cal
L}_{\varphi\varphi} = \overline{\cal L}_{\varphi\varphi}^{int}$
(\ref{r0}), these quantities ${\cal M}$ and ${\cal K}$ are directly
determined in terms of the vacuum expectations $\langle
0|\bm{\varphi}^{2m}(x)|0\rangle$ and
\begin{eqnarray}
\delta_{ab}\Delta_{\mu\nu}=\langle 0|\partial_{\mu}
\varphi_a(x)\partial_{\nu}\varphi_b(x)|0 \rangle =
\delta_{ab} i \int\frac{\mbox{d}^4k}{(2\pi)^4}
\frac{k_{\mu}k_{\nu}}{k^2-m^2+i\varepsilon} , \label{h0}
\end{eqnarray}
which eventually results in
\begin{eqnarray}
{\cal
K}=\frac{2}{3}\sum_{n=1}^{\infty}\frac{(-1)^n(4f)^{2n}}{(2n+2)!}
\langle 0|\bm{\varphi}^{2n}(x)|0\rangle =
\frac{1}{3}\langle
0|\frac{\sin^2y}{y^2}-1|0\rangle , \; \;
y=2f\sqrt{\bm{\varphi}^2(x)} ,
\label{k1} \\
2{\cal M}=m^2\sum_{n=1}^{\infty}\frac{(-1)^{(n+1)}}{3}(2f)^{2n}
\frac{2n+3}{(2n+1)!}\langle
0|\bm{\varphi}^{2n}(x)|0\rangle +\nonumber\\ + \frac{1}{9}
\langle 0|\partial_{\mu}\bm{\varphi}(x)
\partial^{\mu}\bm{\varphi}(x)|0\rangle \sum_{n=1}^{\infty}(-1)^n
(4f)^{2n}\frac{2n(2n+1)}{(2n+2)!}\langle 0|\bm{\varphi}
^{2(n-1)}(x)|0\rangle = \nonumber\\
= m^2\langle
0|1-\frac{1}{3}\Bigl(\cos y + 2\frac{\sin y}{y}\Bigr)|0\rangle
-\frac{2(4f)^2}{9}\langle 0|\partial_{\mu}\bm{\varphi}(x)
\partial^{\mu}\bm{\varphi}(x)|0\rangle \times \nonumber\\
\times\langle 0|\Bigl\{\frac{\sin^2y}{2y^2}+\frac{2}{(2y)^2}\Bigl(\frac
{\sin2y}{2y}-1\Bigr) -\frac{1}{(2y)^2}\Bigl(\frac{\sin^2y}{y^2} - 1
\Bigr)\Bigr\}|0\rangle . \label{m1}
\end{eqnarray}

To proceed, we utilize the directly obtained relations:
\begin{eqnarray}
\langle 0|\bm{\varphi}^{2n}(x)|0\rangle = \Delta^n\cdot(2n+1)!!
\label{h1}
\end{eqnarray}
with
\begin{eqnarray}
\delta_{ab} \Delta = \langle 0|\varphi_a(x)\varphi_b(x)|0\rangle =
\delta_{ab}\int\frac{\mbox{d}^4k}{(2\pi)^4} \frac{i}{k^2-m^2+i\varepsilon} ,
\label{h2}
\end{eqnarray}
and then
\begin{eqnarray}
(2n+1)!! = \frac{2\sqrt{\mathsf p}}{\sqrt{\pi}}\int_{0}^{\infty}(2{\mathsf
p}x^2)^{(n+1)}\exp[-{\mathsf p}x^2] \mbox{d}x , \; \; {\mathsf p}>0 .
\label{h3}
\end{eqnarray}
The integrals (\ref{h2}) and (\ref{h0}) diverge due to the singular
high momenta behavior of the integrand. To interpret them and other
divergent integrals hereafter, the dimensional regularization (or continues dimensional)
method is relevant, see, for instance, refs. \cite{le75,d}.
As ascertained in \hyperlink{appen}{Appendix A}, the quantities (\ref{h2}), (\ref{h0}) are presented in $d$-dimensions in the form
\begin{eqnarray}
\Delta(d)= \frac{\mu^{d-4}}{(4\pi)^{d/2}}\frac{(m^2)^{d/2-1}}{\Gamma(d/2)}\pi(-1)^n
\mbox{cosec}[\pi d/2] , \; \; d/2=n+\nu_d, \; n=0, \pm 1, \pm 2, ... ,  \; \; +0\leq\mbox{Re}\nu_d\leq1-0, \label{d1}\\
\Delta_{\mu\nu}(d) = g_{\mu\nu}\frac{m^2}{d} \Delta(d) , \label{d2}
\end{eqnarray}
where, as usual, $\mu$ is an auxiliary quantity having
dimension of mass, $\Gamma$ is the Euler Gamma-function, and
physical limit corresponds to $d\rightarrow 4, \,  \Delta(d)\rightarrow\Delta(4)\rightarrow +\infty$.

 With the regularization (\ref{d1}), (\ref{d2}), the sequence of summation
and integration can be interchanged in calculating the quantities
${\cal K}$ (\ref{k1}) and ${\cal M}$ (\ref{m1}) with the help of
eq. (\ref{h3}). Then with allowance for eqs. (\ref{h0})
-- (\ref{h3}) we arrive at the eventual results for the quantities
${\cal K , M}$, which determine the pion mass and field
renormalization (\ref{r3}):
\begin{eqnarray}
{\cal K}=\frac{1}{3}\Bigl(P(1-\exp[-1/P])-1\Bigr), \; \; \; \; \; \label{k2}\\
{\cal M}= \frac{m^2}{3}\Bigl(\exp[-1/P] (1+P) - P\Bigr) +
\frac{m^2}{2}\Bigl(1-\exp[-1/4P] + \frac{1}{6P}\exp[-1/4P]\Bigr), \; \; \;
\label{m2}
\end{eqnarray}
where
\begin{eqnarray}
P^{-1} = 2\Delta(2f)^2 . \label{p0}
\end{eqnarray}
The way we have been following here is akin to the well known
methods of divergent series summation, see ref. \cite{hard}.

The main feature to stress is that these quantities ${\cal K , M} ,
{\mathsf z} , m_r^2$ do not diverge, when $d\rightarrow 4$ and
consequently $\Delta\rightarrow\infty , \, P\rightarrow 0$ (see \hyperlink{appen}{Appendix A}). Indeed, in this case we have got
\begin{eqnarray}
{\cal K}=-\frac{1}{3} , \; \; \frac{2{\cal M}}{m^2}=1.
\label{inf}
\end{eqnarray}

Thus, starting with the essentially nonlinear lagrangian
(\ref{2pi}), (\ref{dd}), we encounter no divergence in treating the
mass and wavefunction renormalization. Let us mention that if we
had restricted ourselves by the lowest $\bm{\varphi}^2$ order, we
should apparently have got at $d\rightarrow 4$ the divergent
quantities
\begin{eqnarray}
{\cal K}_{1}=-(2f)^2\frac{\Delta}{3} , \; \; {\mathsf
z}_1^{-1}=1-\frac{2}{3}\Delta (2f)^2 , \; \; \frac{2{\cal
M}_1}{m^2}=\frac{\Delta}{6}(2f) ,  \label{1r1}\\
 m^2_{r \, 1}=m^2\Bigl(1+\frac{\Delta}{2}(2f)^2\Bigr) ,\nonumber
\end{eqnarray}
as discussed in the manifold investigations, see, for instance,
ref. \cite{d}. The method how to deal with this divergence is being
elaborated within the chiral perturbation theory \cite{ap,d}.

The principle feature to stress is that the whole lagrangian
\begin{equation}
{\cal L}_{SBh}={\cal L}_h+{\cal L}_{SB} \, ,\label{hbs}
\end{equation}
 at any values of the parameters involved, is invariant under the
global $SU(2)\times SU(2)$ isospin as well as chiral rotation, save
for the symmetry-breaking term (\ref{sb}). Apparently, due to eqs.
(\ref{1ct}), (\ref{dnp}), any transformation of the field
$\bm{\varphi}$ entails a corresponding transformation of the
nucleon field $N$. The lagrangian (\ref{4l}) proves to be relevant
in treating the low-energy electro-weak processes with non-strange
hadrons, which is our purpose.

\section{Hadron currents}
\label{sec:level3}
The infinitesimal global gauge transformations
\begin{eqnarray}
\delta_{\varepsilon}\psi=-\frac{i}{2}\bm{\tau\varepsilon}\psi \, ,
 \; \; \; \; \delta_{\varepsilon}{\large\bf U}
=-\frac{i}{2}[\bm{\tau\varepsilon},
{\large\bf U}], \; \; \; \;
\delta_{\varepsilon}\varphi_s=-\epsilon_{str}\varphi_t
\varepsilon_r \, , \label{2e}\\
\delta_a\psi=\frac{i}{2}\gamma^5\bm{a\tau}\psi \, , \; \; \; \;
\delta_a{\large\bf U}
=-\frac{i}{2}\{\gamma^5\bm{a\tau},{\large\bf U} \}
\label{2ui}
\end{eqnarray}
of the lagrangian ${\cal L}_{SBh}$ (\ref{hbs}) yield the
iso-triplet of hadronic vector $\bfgr{\cal J}_{\mu}(x)$ and axial
$\bfgr{\cal J}_{5\mu}(x)$ currents$, \,
\{{\cal J}^r_{(5)\mu}(x)\} \, (r=1,2,3)$, one current ${\cal
J}^r_{(5)\mu}(x)$ of the iso-triplet for each $\varepsilon_r , \;
a_r$ in eqs. (\ref{2e}), (\ref{2ui}),
\begin{eqnarray}
\bfgr{\cal J}_{\mu}=\bfgr{\cal J}_{N \, \mu}+
\bfgr{\cal J}_{\varphi \, \mu} \, , \label{tj}\\
\bfgr{\cal J}_{5\mu}=\bfgr{\cal J}_{N \, 5\mu}+
\bfgr{\cal J}_{\varphi \, 5\mu} \, ,\label{5tj}
\end{eqnarray}
where
\begin{eqnarray}
\bfgr{\cal J}_{N \, \mu}=\frac{1}{4}\bar N\gamma^{\mu}
[\xi^{-b}\bm{\tau}\xi^b+\xi^b\bm{\tau}\xi^{-b}]N+
\frac{g'}{4}\bar N\gamma^{\mu}\gamma^5
[\xi^{-b}\bm{\tau}\xi^b-\xi^b\bm{\tau}\xi^{-b}]N=\nonumber\\
=\frac{1}{2}\bar N\gamma^{\mu}
\bm{\tau}N+\bar N
\gamma^{\mu}\Bigl(\bm{\varphi}(\bm{\tau\varphi})
-\bm{\tau}(\bm{\varphi})^2\Bigr)
\frac{\sin^2z}{\bm{\varphi}^2}N+ \; \;  \label{jn} \\
+g'\bar N\gamma^{\mu}
\gamma^5[\bm{\tau}\times\bm{\varphi}]
\frac{\sin2z}{2\sqrt{\bm{\varphi}^2}}N\approx\bar
N\gamma^{\mu}
\frac{\bm{\tau}}{2}N + {\cal O}(f^2) \, , \nonumber\\
\bfgr{\cal J}_{N \, 5\mu}=\frac{g'}{4}\bar N \gamma^{\mu}\gamma^5
[\xi^{-b}\bm{\tau}\xi^b+\xi^b\bm{\tau}\xi^{-b}]N+\frac{1}{4}
\bar N \gamma^{\mu}
[\xi^{-b}\bm{\tau}\xi^b-\xi^b\bm{\tau}\xi^{-b}]N=\nonumber \\
=\frac{g'}{2}\bar N\gamma^{\mu}\gamma^5\bm{\tau}N+
g'\bar N \gamma^{\mu}\gamma^5
\Bigl(\bm{\varphi}(\bm{\tau\varphi})
-\bm{\tau}(\bm{\varphi})^2\Bigr)
\frac{\sin^2z}{\bm{\varphi}^2}N+ \label{5jn}\\
+\bar N\gamma^{\mu}
[\bm{\tau}\times\bm{\varphi}]
\frac{\sin2z}{2\sqrt{\bm{\varphi}^2}}
N\approx g'\bar N\gamma^{\mu}\gamma^5
\frac{\bm{\tau}}{2}N + {\cal O}(f^2)\, , \nonumber
\end{eqnarray}
\begin{eqnarray}
\bfgr{\cal J}_{\varphi \, \mu}=-\frac{i}{16f^2}\mbox{Sp}[\bm{\tau}
\bigl({\large\bf U}^{\dagger}_{\xi}\partial_{\mu}
{\large\bf U}_{\xi}+ {\large\bf U}_{\xi}\partial_{\mu}{\large\bf
U}^{\dagger}_{\xi}\bigr)]= \label{jp}\\
=\Bigl(\frac{\sin y}{y}\Bigr)^2
[\bm{\varphi}\times\partial_{\mu}\bm
{\varphi}]\approx[\bm{\varphi}\times\partial_{\mu}\bm
{\varphi}]+{\cal O}(f^3) \, ,
\nonumber\\
\bfgr{\cal J}_{\varphi \, 5\mu}=
\frac{-i}{16f^2}\mbox{Sp}[\bm{\tau}
\bigl({\large\bf U}^{\dagger}\partial_{\mu}
{\large\bf U}-{\large\bf U}
\partial_{\mu}{\large\bf U}^{\dagger}\bigr)\gamma^5]
=\frac{-i}{16f^2}\mbox{Sp}[\bm{\tau}
\bigl({\large\bf U}^{\dagger}_{\xi}\partial_{\mu}
{\large\bf U}_{\xi}-{\large\bf U}
_{\xi}\partial_{\mu}{\large\bf U}^{\dagger}_{\xi}\bigr)]=
\label{5jp}\\
=\frac{\partial_{\mu}\bm{\varphi}}{2f}
\Bigl(\frac{\sin 2y} {2y}\Bigr)+
f \frac{2\bm{\varphi}\partial_{\mu}\bm{\varphi}^2}
{y^2}\Bigl(1-\frac{\sin 2y}{2y}\Bigr) \approx\frac{1}{2f}
\partial_{\mu}\bm{\varphi}+{\cal O}(f) \, .\nonumber
\end{eqnarray}

The vector current is conserved,
\begin{eqnarray}
\partial^{\mu}\bfgr{\cal J}_{\mu} =0 \, . \label{td}
\end{eqnarray}
Yet as the total lagrangian ${\cal L}_{SBh}$ (\ref{hbs}) does
 besides ${\cal L}_h$ include the symmetry-breaking term ${\cal
L}_{SB}$, the relation holds
\begin{eqnarray}
\partial^{\mu}\bfgr{\cal J}_{5\mu}
=\frac{i m^2}{32f^2}\mbox{Sp}\{({\large\bf U}_{\xi}^{\dagger}-
{\large\bf
U}_{\xi}),\bm{\tau}\}=\bm{\varphi}\frac{m^2}{2f}\frac {\sin
y}{y}\approx\frac{m^2}{2f}\bm{\varphi}+{\cal O}(f) \, ,
\label{5td}
\end{eqnarray}
the axial current is partially-conserved. It is to stress that the
eqs. (\ref{td}), (\ref{5td}) hold true just for the total currents,
yet not for the nucleon (\ref{jn}), (\ref{5jn}) and the pion
(\ref{jp}), (\ref{5jp}) currents themselves separately, so far the
nucleon-pion interactions are involved into
 ${\cal L}_h$.

  The compensating field $\bm{\varphi}(x)$ is associated with the
physical pion field,
\begin{equation}
\pi^{\pm}(x)=\frac{1}{\sqrt{2}}[\varphi_1(x)\mp i\varphi_2(x)]
 \, , \; \; \pi^0(x)=\varphi_3(x) \, . \label{pf}
\end{equation}
The lagrangian ${\cal L}_{SBh}$ (\ref{hbs}) and the currents
(\ref{tj}), (\ref{5tj}) are expressed in terms of the iso-doublet
of nucleons, the neutron and proton, $N=n, p$, and the iso-triplet
of pions, $\pi^r , \, r=\pm , \, 0$.

 The lagrangian ${\cal L}_{SBh}(x)$ (\ref{hbs}) of considered
system is also invariant under the global $U_Y(1)$ transformation,
with the hypercharge $Y=1$ for the nucleon iso-doublet and $Y=0$
for the pion iso-triplet. Then invariance of ${\cal L}_{SBh}(x)$
under the $U_Y(1)$ transformation of nucleon fields,
\begin{eqnarray}
\psi'_N(x)\Longrightarrow\psi_N(x)+\delta_{0}\psi(x) \, , \; \;
\delta_{0}\psi(x)=-i\varepsilon_0{\bf
I}_{NN'}\psi_{N'}(x)\frac{Y}{2} \, , \; \; \; \; Y=1 \, , \label{y}
\end{eqnarray}
 yields the conserved neutral iso-scalar current
\begin{eqnarray}
 {\cal J}^0_{\mu}(x)=\frac{1}{2}{\bf I}_{NN'}(\bar
 N\gamma^{\mu}N'). \label{i12}
\end{eqnarray}
Certainly, any linear combination of the conserved (or
partially-conserved) currents (\ref{tj})-(\ref{5jp}), (\ref{i12})
is a conserved (or partially-conserved) current as well.

 The combination
\begin{eqnarray} J^0_{\mu}(x)\equiv J^{em}_{\mu}(x)= {\mathsf
Z}_e{\cal J}^3_{\mu}(x)+{\cal J}^0_{\mu}(x)
\, ,
\label{0c}
\end{eqnarray}
of the third component of the iso-vector current (\ref{tj}) and the
iso-scalar current (\ref{i12}), is naturally understood to be the
electromagnetic current, which stands to describe interaction of a
hadronic system with an electromagnetic field. So, hereafter any
current with the upper index $0$ is implied to be the
electromagnetic current, $J^{0}_{\mu}(x)\equiv J^{em}_{\mu}(x)$.

The expectation value of electromagnetic current in the state
$|N(p),0\rangle$ containing no pions (pion vacuum) and a single
nucleon, the proton or neutron, with a momentum $p$ is obtained
(with the proper wave functions normalization) from eqs.
(\ref{jn}), (\ref{i12})
\begin{eqnarray}
\int \langle N, 0\mid J^{em}_{\mu}(x)\mid N , 0\rangle
\mbox{d}{\bf x}=\nonumber\\
=\chi^{*}_N\frac{p_{\mu}}{p_0}[\frac{1}{2}({\mathsf Z}_e\tau_0 + {\bf
I})-{\mathsf Z}_e\tau_0\frac{2}{3}\langle 0\mid \sin^2z \mid
0\rangle]\chi_N, \; \; (z=f\sqrt{\bm{\varphi}^2(x)}) \label{1j}
\end{eqnarray}
where $\chi_N$ stands for nucleon isospinor. To this expression at
$\mu=0$ reduce to zero for the neutron $(N=n)$ and to unite for the
proton $(N=p)$, as required, there should be
\begin{eqnarray}
{\mathsf Z}_e=\Bigl(1-\frac{4}{3}\langle 0\mid\sin^2z\mid
0\rangle\Bigr)^{-1}
 . \label{1j1}
\end{eqnarray}
Here the vacuum expectation is evaluated just likewise the akin
quantities in the expressions (\ref{k1})-(\ref{h3}), so that we
have got
\begin{eqnarray}
\langle 0\mid\sin^2z\mid 0\rangle = \frac{1}{2}-\frac{2 P -1}{4
P}\exp[-\frac{1}{4P}] , \label{s}
\end{eqnarray}
with $P$ given by eq. (\ref{p0}). Of course, in the lowest
$\bm{\varphi}^2$-order there would be ${\mathsf Z}_e=1$. At the
physical limit, $d\rightarrow 4 , \; \Delta\rightarrow\infty \; $
in eqs. (\ref{d1}), (\ref{d2}) (see \hyperlink{appen}{Appendix A}), this reduces to
\begin{equation}
\langle 0\mid\sin^2z\mid 0\rangle = \frac{1}{2} , \label{s1}
\end{equation}
 and we have got ${\mathsf Z}_e =3$, in treating the essentially
 non-linear current (\ref{jn}).

  In the usual way, the first and second components of isovector
current ${\cal J}^{r=1,2}_{\mu}(x)$ serve to construct the charge
vector currents
\begin{eqnarray}
J^{\pm}_{N \mu}(x)= {\mathsf Z}_e({\cal J}^1_{N \mu}(x)
\pm i{\cal J}^2_{N \mu}(x)) \, , \label{pmc}
\end{eqnarray}
with the same parameter $ {\mathsf Z}_e $ as in eq. (\ref{0c}) where it
occurred as a factor by the third component of isovector current.
 As one observes, the current $J^{+}_{(5)\mu}(x)$ would increase
 whereas the current $J^{-}_{(5)\mu}(x)$ decrease electric charge
 of a hadron system by unity. The currents
  $J^{\pm}_{(5)\mu}(x)$ occur in treating the $\beta-$decay of
hadrons. For instance, the vector part of $n
\rightarrow p$ weak transition current is the matrix element of
$J^{+}_{N \mu}(x)$ (\ref{pmc}) between single neutron and single
proton states (and pion vacuum)
\begin{eqnarray}
\langle p,0\mid J^{+}_{N\mu}\mid n,0\rangle = {\mathsf Z}_e \bar
U_p\gamma_{\mu}\Bigl(1-\frac{4}{3}\langle 0\mid\sin^2z(x)\mid
0\rangle \Bigr)U_n , \label{j3}
\end{eqnarray}
where $\bar U_p , \, U_n$ are the proton and neutron wave
amplitudes. Thereby one infers the parameter ${\mathsf Z}_e$ emerges
again to be defined by the expression (\ref{1j1}).

Calculating the axial part of the respective transition current
serves to express the parameter $g'$ in the lagrangian (\ref{il1})
through the nucleon axial form-factor $g_A$:
\begin{eqnarray}
\langle p,0\mid J^{+}_{N 5 \mu}(x)\mid n,0\rangle
= \langle p,0\mid {\cal J}^1_{N 5 \mu}(x) + i{\cal J}^2_{N 5
\mu}(x)\mid n,0\rangle = \nonumber\\
=g'\bar U_p\gamma_{\mu}\gamma_5\Bigl(1-\frac{4}{3}\langle 0
\mid \sin^2z(x)\mid 0\rangle U_n =g_A \bar U_p\gamma_{\mu}\gamma^5
U_n \, , \label{j4}
\end{eqnarray}
with the vacuum expectation value given by eq. (\ref{s}). Thus we
have got
\begin{eqnarray}
g'=g_A\Bigl\{1-\frac{1}{3}\Bigl
(2-\frac{2P-1}{P}\exp[-{1}{/}{4P}]\Bigr)\Bigr\}^{-1} , \label{j5}
\end{eqnarray}
which reduces to
\begin{equation}
g'=3g_A ,\label{j6}
\end{equation}
 for the physical limit, $d\rightarrow 4,\, \Delta\rightarrow\infty,\, P\rightarrow 0$ (see \hyperlink{appen}{Appendix A}), of
eqs. (\ref{d1}), (\ref{p0}). Of course, in the lowest
$\bm{\varphi}^2$-order ( $P^{-1}\rightarrow 0$ ) there would
merely be $g'=g_A$.

 With having recourse to the Standard Model (SM) concepts (see, for
instance, \cite{d}), the third component ${\cal J}^3_{(5)\mu}(x)$
of the isotriplet current $\bfgr{\cal J}_{(5)\mu}(x)$ and the
iso-scalar current ${\cal J}^0_{\mu}$ are still combined giving,
besides $J^{em}_{\mu}(x)$, the neutral weak currents
\begin{eqnarray}
J^Z_{\mu}(x)={\mathsf Z}_e {\cal J}^3_{\mu}(x) \bigl(1-2 s^2_W
\bigr)-2{\cal J}^0_{\mu}(x) s^2_W \, , \; \;
s^2_W=1-\frac{M^2_W}{M^2_Z} \, ,
\label{zc}  \\
J^Z_{5\mu}(x)= {\cal J}^3_{5\mu} \; . \label{z12}
\end{eqnarray}
Here the coefficients are chosen so that these currents (\ref{zc}),
(\ref{z12})
 would serve to describe the feasible hadron transitions in the
neutral $Z^0$-boson field (see, for instance, \cite{d}).

\section{The hadron interaction with gauge fields.}
\label{sec:level4}
The hadron currents have been considered, we are now to describe
the hadron (pion and nucleon) interactions with the electromagnetic
field ${\cal A}^{em}_{\mu}(x)$, and with the charged ${\cal
A}_{\mu}^{\pm}(x)$ and neutral ${\cal A}^Z_{\mu}(x)$ fields
associated with the $W^{\pm}-$ and $Z^0-$boson fields. The
electro-weak lagrangian ${\cal L}^{EW}_{int}(x)$ to describe these
interactions originates from requirement of ${\cal L}_h$ (\ref{4l})
invariance under the local ({\it i.e.} with the space-time dependent
parameters $\bm{\varepsilon}(x) ,
\; {\bf a}(x)$) gauge transformation (\ref{2e}), (\ref{2ui}),
(\ref{y}).

In order for the local transformation (\ref{2e}), (\ref{2ui}),
(\ref{y}) to be an invariance of the lagrangian ${\cal L}_h(x)$
(\ref{4l}), the derivatives $\partial_{\mu}\psi_N(x) , \;
\partial_{\mu}{\large\bf U}(x)$ in ${\cal L}_h(x)$ are known (see,
 for instance, \cite{d,ll,ad,af}) to be replaced by the respective
covariant derivatives ${\cal D}_{\mu}(x)\psi_N(x) , \; {\cal
D}_{\mu}(x){\large\bf U}(x)$, which read for our consideration as
\begin{eqnarray}
\partial_{\mu}\psi_N(x)\Longrightarrow{\cal D}_{\mu}(x)\psi_N(x) =
\partial_{\mu}\psi_N(x) + i e {\cal A}_{\mu}^a(x)
\ell^a_{NN'}\psi_{N'}(x) , \; \; \; e=\sqrt{4\pi\alpha} ,
 \, \label{d}
\end{eqnarray}
where
\begin{eqnarray}
\ell^a_{NN'}=\frac{1}{2}\delta_{a0}({\bf
I}_{NN'}+{\mathsf Z}_e\tau^0_{NN'})+a^2\tau^a_{NN'}({\mathsf Z}_e-\gamma^5)
, \; \; a=0,\pm , \label{d0}\\
\tau^0=\tau_3 , \; \; \; \tau^{\pm}=(\tau_1 \pm
i\tau_2)/2 , \nonumber\\
\ell^Z_{NN'}=-{\bf
I}_{NN'}s_W^2+\frac{1}{2}{\mathsf Z}_e\tau^0_{NN'}(1-2s_W^2)-
\gamma^5\frac{1}{2}\tau^0_{NN'} , \; \; a=Z , \label{dz}
\end{eqnarray}
and
\begin{eqnarray}
\partial_{\mu}{\large\bf U}(x)\Longrightarrow{\cal D}_{\mu}(x)
{\large\bf U} (x)=\partial_{\mu}{\large\bf U}(x)+\frac{i}{2}e{\cal
A}_{\mu}^0(x)[\tau^0,{\large\bf U}(x)]+
\nonumber\\
+i e {\cal A}_{\mu}^{\pm}(x)\Bigl([\tau^{\pm},{\large\bf
U}(x)]+\{\tau^{\pm}
\gamma^5,{\large\bf U}(x)\}\Bigr)+ \label{du}\\
+\frac{i e}{2}{\cal A}^Z_{\mu}(x)\Bigl([\tau^0,{\large\bf
U}(x)](1-2s_W^2)+
\{\tau^0\gamma^5,{\large\bf U}(x)\}\Bigr). \nonumber
\end{eqnarray}
 The parameter ${\mathsf Z}_e$ (\ref{1j1}) involved herein has been
determined so that to provide the proper description of hadron
neutral and charge currents. Then the electroweak lagrangian ${\cal
L}_{int}^{EW}(x)$ results in terms of the hadron currents
$J^a_{\mu}(x)$ (\ref{tj})-(\ref{z12}) and the fields ${\cal
A}^a_{\mu}(x)$, which are related to the physical fields $
{A}_{\mu}^{em}(x)\, ,\; W^{\pm}_{\mu}(x) \, , \; Z^0_{\mu}(x)$, and
the effective coupling constants by the equations
\begin{eqnarray}
{\cal A}_{\mu}^0(x)={A}_{\mu}^{em}(x) \; , \; \; e{\cal
A}_{\mu}^{\pm}(x)=
\frac{\mid V_{ud}\mid\sqrt{G}M_W}{2^{{1}{/}{4}}}W_{\mu}^{\pm} \, ,
\; \; e{\cal A}_{\mu}^{Z}(x) =
\frac{\mid V_{ud}\mid\sqrt{2G}M_Z}{2^{{1}{/}{4}}}Z^0_{\mu} \, , \qquad
\label{aa} \\
\frac{\sqrt{G}}{2^{{1}{/}{4}}}=\frac{|e|}{2\sqrt{2}s_W M_W}
 , \; \; \; \; \; \; \; \; \; \; \; \qquad \nonumber
\end{eqnarray}
with the Fermi interaction constant $G$, the $W^{\pm}-$ and $Z^0-$
boson masses $M_W , M_Z$, and the $CKM-$quark-mixing matrix element
$|V_{ud}|$ \cite{pdg}, so that to provide true description of
hadron interactions with the electromagnetic and heavy gauge boson
fields \cite{d,bb}.

Then we arrive at the effective lagrangian to explore the pion and
nucleon interactions with the gauge fields,
\begin{eqnarray}
 {\cal L}^{EW}_{int}(x) = L^{em}(x ) +
  L^{1W}(x) + L^{Wem}(x) + L^{1Z}(x) + L^{Zem}(x) \, ,
 \qquad\label{ew}\\
   L^{em}(x) = L^{1em}(x) + L^{2em}(x),
\qquad\label{em}\\
   L^{1em}(x) = -e {\cal A}_{\mu}^0(x) J^{0 \, \mu}(x) = -e
   {A}_{\mu}^{em}(x) J^{em \, \mu}(x) \; ,
\qquad\label{1em}\\
  L^{2em}(x) = e^2 {\cal A}_{\mu}^{0} {\cal A}^{0 \, \mu} \pi^-(x)
 \pi^+(x)\frac{\sin^2y}{y^2} = e^2 A^{em}_{\mu}A^{em \, \mu}
\frac{\sin^2y}{y^2} \pi^-(x) \pi^+(x), \qquad \label{2em}\\
  L^{1W}(x) = - e {\cal A}^{a \, \mu}(x) {\tilde J}^a_{\mu}(x) =
- \frac{\mid V_{ud}\mid\sqrt{G}M_W}
{2^{{1}{/}{4}}} {\tilde J}^a_{\mu}(x) W^{a \,
\mu}(x), \; \; \; a=\pm ,
\qquad \label{1w}\\
  L^{Wem}(x) = -eA_{\mu}^{em}(x)
\frac{\mid V_{ud}\mid\sqrt{G}M_W}{2^{{1}{/}{4}}}
W^{a \, \mu}(x)\sqrt{2}\pi^{-a}
\Bigl(\pi^0\frac{\sin^2(y)}{y^2}+a\frac{i}{2f}\frac{\sin(2y)}{2y}
\Bigr) , \qquad\label{wem}\\
  L^{1Z}(x)=-
\frac{\mid V_{ud}\mid\sqrt{2G}M_Z}{2^{{1}{/}{4}}}
{\tilde J}^Z_{\mu}(x) Z^{0 \, \mu}(x),
\qquad\label{1z}\\
 L^{Zem}(x)=eA_{\mu}^{em}(x)\frac{\mid
 V_{ud}\mid\sqrt{2G}M_Z}{2^{{1}{/}{4}}} Z^{0 \,
 \mu}(x)\frac{\sin^2(y)}{y^2}(1-2s^2_W)\pi^{+}(x)\pi^{-}(x)
  \, .\qquad\label{zem}
\end{eqnarray}
All the interactions quadratic in the fields ${\cal
A}^{\pm}_{\mu}(x) \, , \; {\cal A}^Z_{\mu}(x) $ are abandoned in
${\cal L}^{EW}(x)$ (\ref{ew}), yet the interactions $L^{2em} $
(\ref{2em}) , $ \, L^{Wem}$ (\ref{wem}) , $\, L^{Zem} $ (\ref{zem})
, incorporating the products of fields, $A_{\mu}^{em} A^{em \mu} ,
\, A_{\mu}^{em} W^{a
em} , \, A_{\mu}^{em} Z^{0\,\mu} $, are presented therein. Thus, the
lagrangian ${\cal L}^{EW}_{int}$ enables one to describe the
electro-weak transitions accompanied immediately by
$\gamma-$radiation. The expressions (\ref{1em}), (\ref{1w}),
(\ref{1z}) are written in terms of the currents
\begin{eqnarray}
J^0_{\mu}(x)\equiv J^{em}_{\mu}(x) \, , \; \; {\tilde
J}_{\mu}^{\pm}(x)=J_{\mu}^{\pm}(x)+J_{5 \, \mu}^{\pm} \, ,
\; \; \; {\tilde J}_{\mu}^{Z}(x)=J_{\mu}^Z(x)+J_{5 \, \mu}^Z ,
\label{+5}
\end{eqnarray}
where the currents $J_{(5) \, \mu}^{0 \, , \pm \, , \, Z}$ are
derived in Sec. III. As realized, the emission (absorption) of
$W^{\pm}-,Z^0-$bosons by hadrons is described by the lagrangian
(\ref{ew}).

The total lagrangian $\tilde{\cal L}(x)$ to describe the hadron
electro-weak transitions in considered system is the sum of the
lagrangians
 ${\cal L}_{SBh}(x)$ (\ref{hbs}) and ${\cal L}_{int}^{EW}(x)$
 (\ref{ew}),
\begin{eqnarray}
\tilde{\cal L}(x)={\cal L}_{SBh}(x)+{\cal L}_{int}^{EW}(x) .
\label{+l}
\end{eqnarray}

The lagrangian (\ref{ew}), (\ref{+l}) is to be complemented by the
ordinary lagrangian ${\Large\ell}^{EW}(x)$ to describe the
interactions of leptons with photons and heavy gauge bosons :
\begin{eqnarray}
{\Large\ell}^{EW}(x) = {\Large\ell}^{em}(x) + {\Large\ell}^{1W}(x)
+ {\Large\ell}^{1Z}(x) , \label{wl}\\ {\Large\ell}^{em}(x) =-
e\bar\psi_l(x)\gamma^a\psi_l(x) A^{em}_a(x) , \; \; l=e, \mu , \tau
, \label{wll}\\ {\Large\ell}^{1W}(x)=-\frac{M_W\sqrt{G}}{2^{1/4}}
\Bigl(W^{+}_a(x)\bar\psi_{\nu}(x)\gamma^a (1+\gamma^5)\psi_l(x)
+\nonumber\\
 +W^{-}_a(x)\bar\psi_l(x)\gamma^a (1+\gamma^5)\psi_{\nu}(x)
\Bigr) ,\label{wlw}\\ {\Large\ell}^{1Z}(x)=-\frac{M_Z
\sqrt{2G}}{2^{1/4}}Z^0
\Bigl(\frac{1}{2}\bar\psi_{\nu}(x)\gamma^a
(1+\gamma^5)\psi_{\nu}(x) +\nonumber\\+ \bar\psi_l(x)\gamma^a
(-\frac{1}{2}(1+\gamma^5)+2s^2_W)\psi_l(x)\Bigr) , \label{wlz}
\end{eqnarray}
were $\psi_{\nu} , \, \psi_{l=e,\mu ,\tau}$ stand for the neutrino,
 and electron, muon, $\tau-$lepton fields.

 The contribution ${\cal L}_{gauge}({\cal A}^a,\partial_{\mu}{\cal
A}^a)$ to the total lagrangian from the interactions of the gauge
fields ${\cal A}^a_{\mu}(x)$ themselves is not immediately engaged
into consideration of the low-energy electro-weak transitions of
hadrons. So, there is here no need to plunge into its construction
and add ${\cal L}_{gauge}({\cal A}^a,\partial_{\mu}{\cal A}^a)$ to
the effective lagrangian of the considered system.

The lagrangians we have been treating incorporate a number of
parameters which are specified by correlating the evaluated
physical quantities with the appropriate experimental data.

Physical content and magnitude of the primary parameter $f$ is
acquired by considering the pion weak decay,
\begin{eqnarray}
\pi^{+}\Longrightarrow \mu^{+} \nu_{\mu} \; (e^{+}\nu_{e}) .
\label{pmn}
\end{eqnarray}

The interaction
\begin{eqnarray}
L^{\pi}(x) = {\Large\ell}^{1W}(x) + L^{1W}(x) \label{0d}
\end{eqnarray}
causing this process is read of from the expressions (\ref{1w}),
(\ref{+5}), (\ref{wlw}). With proper allowance for eqs.
(\ref{5jp}), (\ref{aa}), (\ref{h1})--(\ref{d1}), the
$S_{\pi}-matrix$ element to describe the transition (\ref{pmn}) is
given by
\begin{eqnarray}
\langle \nu , \mu^+ |S_{\pi}|\pi^{+} \rangle
= -\frac{1}{2}\langle \nu , \mu^+
|{\cal T}\int\mbox{d}^4x_1 \int\mbox{d}^4x_2 L^{\pi}(x_1)
L^{\pi}(x_2)|\pi^{+}\rangle=\nonumber\\
=-\langle \nu , \mu^+ |{\cal T} \int\mbox{d}^4x_1 \int\mbox{d}^4x_2
J^{-}_{\varphi 5 \, \alpha}(x_1) W^{- \, \alpha}(x_1)\times
\nonumber\\
\times\bar\psi_{\nu}(x_2)\gamma^{\beta}
(1+\gamma^5)\psi_{\mu}(x_2) W_{\beta}^{+}(x_2)
|\pi^{+}\rangle\frac{G|V_{ud}|M_W^2}{\sqrt{2}} = \label{1d}\\
 = (2\pi)^4\delta (P_{\pi} - p_{\nu} -
 p_{\mu})\frac{G|V_{ud}|}{2\tilde f}u_{\pi}P_{\pi}^{\alpha} \cdot
 (\bar{u}_{\nu} \gamma_{\alpha} (1+\gamma^5 )u_{\mu} ) , \nonumber
\end{eqnarray}
where the renormalized quantity $\tilde{f}$, associated now with
the pion decay (\ref{pmn}) vertex, appears in place of the initial,
non-renormalized parameter $f$:
\begin{eqnarray}
\frac{1}{2\tilde f} = \frac{1}{2
f}\Bigl(1+\frac{1}{3}\sum_{n=1}^{\infty}\frac{(-1)^n(4f)^{2n}}
{(2n+1)!}\langle 0|\bm{\varphi}^{2n} |0\rangle\Bigr)
=\frac{1}{2f}\Bigl(1+\frac{1}{3}\langle 0|\frac{\sin
2y}{2y}-1|0\rangle \Bigr)=\nonumber\\
=\frac{1}{2f}\frac{1}{3}\Bigl(2+\exp[-1/P]\Bigr)
 , \label{f1}
\end{eqnarray}
with $P$ given by eq. (\ref{p0}).
 In the expression (\ref{1d}), $P_{\pi}, u_{\pi} , \; p_{\nu},
 u_{\nu} , \; p_{\mu}, u_{\mu}$ stand for momenta and wave
amplitudes of the pion, neutrino and muon respectively. Needless to
say all these momenta are negligible as compared to the gauge boson
 mass $\mbox{M}_W$, that appears in calculating the quantity
 (\ref{1d}) through the $W$-boson propagator. By way of
illustration, the amplitude (\ref{1d}), as caused by the
interaction (\ref{0d}), can be displayed by the diagram, \hyperlink{ris1}{fig. 1.} ,
where the small disk stands for $W$ coupling with leptons, and the
double-circle figures pion-$W$ coupling, with all the infinite
number of virtual pion propagators. The solid line depicts a
decaying pion, and the zigzag one displays a $W$-boson propagator;
the thin lines stand for final leptons. (As $M_W\rightarrow\infty$,
the zigzag gets constricted in point, $x_1\rightarrow x_2$.)

The transition amplitude (\ref{1d}) is still modified by the pion
wave function renormalization
\begin{eqnarray}
u_{\pi}=u_{\pi \, r}{\mathsf z}^{1/2}=u_{\pi r} (1-{\cal K}),
\label{uz}
\end{eqnarray}
according to Sec. 2, eq. (\ref{r3}). That results in the additional
renormalization of the amplitude (\ref{1d}) by replacing
\begin{eqnarray}
\frac{1}{2\tilde f}\Longrightarrow\frac{1}{2f_r}=
\frac{1}{2\tilde
f}\Bigl(1-\frac{1}{3}\Bigl(P\Bigl(1-\exp[-{1}{/P}]\Bigr)-1\Bigr)
\Bigr)=\nonumber\\
=\frac{1}{2f}\Bigl\{1+\frac{1}{3}\Bigl(\exp[-{1}{/}{P}]-1\Bigr)-
\frac{1}{3}\Bigl(P\Bigl(1-\exp[-{1}{/}{P}]\Bigr)-1\Bigr)\Bigr\},
\label{f2}
\end{eqnarray}
which is to be equated with the pion decay constant,
\begin{equation}
\frac{1}{2f_r}=F_{\pi}\approx 92.3 \mbox{MeV}. \label{jj}
\end{equation}
 The point to emphasize is that no divergence occurs in calculating
the renormalized decay amplitude (\ref{1d}), (\ref{f2}), just alike
in calculating the quantities (\ref{r3}), (\ref{1j}), (\ref{j3}),
albeit the vacuum expectation (\ref{d1})
 itself diverges at the physical limit: $\Delta\rightarrow\infty , \,  P\rightarrow 0$ ,
when $ d\rightarrow 4 $ (see \hyperlink{appen}{Appendix A}). In this case one obtains from eq.
(\ref{f2}) the plain relation
\begin{eqnarray}
\frac{1}{2f}=\frac{1}{2f_r}=F_{\pi}
\label{mf}
\end{eqnarray}
 to express the parameter $f$, that resides in the effective
lagrangian, through the phenomenological pion decay constant
$F_{\pi}$.

It is expedient to correlate the parameters $g' , \, g_A , \, f ,
\, F_{\pi} $ residing in the lagrangian (\ref{4l}) with the
pion-nucleon amplitude $g_{NN\pi}$ which would determine the
simplest Lorenz-invariant pion-nucleon interaction,
\begin{eqnarray}
{\cal L}_{NN\pi}=ig_{NN\pi} {\bar N} \gamma_5 (\bm{\tau \varphi})
 N \, .\label{ja}
\end{eqnarray}
According to the interaction ${\cal L}_{1 N\varphi}^{int}$
(\ref{il1}), we obtain the amplitude of the charge (negative) pion
emission process $n\rightarrow p+\pi^-$ (the pion-nucleon vertex)
as follows
\begin{eqnarray}
\langle p,\pi^-\mid S_{np\pi}\mid n,0\rangle =
-i g' \langle p,\pi^-\mid \int \mbox{d}^4x \bar
N\gamma^{\mu}\gamma^5\times\nonumber\\
\times
\Bigl\{\frac{\bm{\tau}\partial_{\mu}\bm{\varphi}}{2\sqrt{\bm
{\varphi}^2}}\sin 2z +
(\bm{\tau\varphi})(\bm{\varphi}\partial_{\mu}\bm{\varphi})f
\frac{1}{\bm{\varphi}^2}\Bigl(1-\frac{\sin 2z)}{2z}\Bigr)
\Bigr\} N \mid n , 0\rangle=\nonumber\\
{=}g'f\sqrt{2}(2\pi)^4\delta(p_n-p_p-q_{\pi})\nonumber\\
\Bigl\{\frac
{1}{3}\Bigl(1{-}\exp[-{1}{/}{4P}]\Bigr){+}\exp[-{1}{/}{4P}]\Bigr\}
u_{\pi r}{\bar U}_p\gamma_{\mu}\gamma^5 q^{\mu}_{\pi} U_n ,
\label{jb}
\end{eqnarray}
where $p_n , \, p_p , \, q_{\pi} $ are neutron, proton and pion
(off-mass-shell) momenta, and we have evaluated the expectation
values as we did thus far, $P$ being defined by eq. (\ref{p0}). The
parameter $g'$ is related to the nucleon axial form-factor $g_A$ by
 eqs. (\ref{j5}), (\ref{j6}), and $f$ is given in terms of the pion
decay constant $F_{\pi}$ according to eqs. (\ref{f2})-(\ref{mf}).
In the physical limit, $d\rightarrow 4 , \, P\rightarrow 0 $ (see \hyperlink{appen}{Appendix A})\,
 , \, and $g'\rightarrow 3g_A , \,$ this eq.(\ref{jb}) reduces to
\begin{eqnarray}
\langle p,\pi^-\mid S_{np\pi}\mid n,0\rangle{=}
g_A f\sqrt{2}(2\pi)^4\delta(p_n-p_p-q_{\pi}) u_{\pi r}{\bar
U}_p\gamma_{\mu}\gamma^5 q^{\mu}_{\pi} U_n .
\label{jb2}
\end{eqnarray}
Using the Dirac equation for on-mass-shell nucleons, the expression
(\ref{jb}) transforms to
\begin{eqnarray}
\langle p,\pi^-\mid S_{np\pi}\mid n,0\rangle =
-g'(M_n+M_p) f\sqrt{2} (2\pi)^4\delta(p_n-p_p-q_{\pi})\times
\nonumber\\
\times\Bigl\{\frac
{1}{3}\Bigl(1{-}\exp[-{1}{/}{4P}]\Bigr){+}\exp[-{1}{/}{4P}]\Bigr\}
u_{\pi r}{\bar U}_p\gamma^5 U_n . \label{mpn}
\end{eqnarray}

On the other hand, with describing $N\pi-$interaction by eq.
(\ref{ja}), the emission amplitude of a pion with the same wave
amplitude $u_{\pi r}$ proves to be
\begin{eqnarray}
\langle p,\pi^-\mid S_{np\pi}\mid n,0\rangle =
 -g_{NN\pi}\langle p,\pi^-\mid\int{\bar
N}\gamma^5(\bm{\tau\varphi})N \mbox{d}^4x \mid n,0\rangle
=\nonumber\\
= -g_{NN\pi}\sqrt{2}(2\pi)^4\delta(p_n-p_p-q_{\pi}) u_{\pi r}
{\bar U}_p\gamma^5 U_n. \label{jc}
\end{eqnarray}
From the eqs. (\ref{jc}) and (\ref{mpn}) one infers
\begin{eqnarray}
g_{NN\pi}= g' (M_p+M_n) f\Bigl\{\frac
{1}{3}\Bigl(1-\exp[-{1}{/}{4P}]\Bigr)+\exp[-{1}{/}{4P}]\Bigr\}.
 \label{jd}
\end{eqnarray}
At the physical limit, that is at $d\rightarrow 4 , \,
\Delta\rightarrow\infty , \, P\rightarrow 0$ (see \hyperlink{appen}{Appendix A}) in the expressions
(\ref{d1}), (\ref{p0}) and $g'=3g_A$ according to eq. (\ref{j6}),
this reduces to
\begin{eqnarray}
g_{NN\pi}=\frac{(M_p + M_n)g_A}{2 F_{\pi}}, \label{je}
\end{eqnarray} which is generally referred to
as the Goldberger-Teiman identity \cite{gtr}. So, in treating the
essentially nonlinear pion-nucleon lagrangian, this well known
relation is valid. Thus, all the parameters involved in our
treatment have been specified.

\section{On the pion-nucleon weak interaction.}
\label{sec:level5}

The developed lagrangian ${\cal L}^{EW}_{int}$ (\ref{ew}) describes
the hadron interaction with the gauge fields. This enables one to
consider the weak hadron-hadron interaction, which is due to the
$W^{\pm}- , \, Z^0 -$bosons exchange. The $P$-invariance violation
in pion-nucleon interactions is a generally received manifestation
of such processes \cite{sha,DDH,DZ,Hol}. The hadron
 weak-interaction physics is comprised in the parity-violating
$NN\pi-$vertex. The charge (negative) pion emission $n\rightarrow
p+\pi^-$, that was discussed in Sec. 4, typifies these phenomena.
Side by side with the expression (\ref{jb}), the total amplitude of
this reaction (the total vertex) comprises a part caused by the
interaction $L^{1W}$ (\ref{1w}) which violates $P-$parity,
\begin{eqnarray}
\langle{p,\pi^-}\mid S_{np\pi}^W\mid{n,0}\rangle
{=}-\frac{1}{2}\langle p,\pi^-\mid{\cal
T}\int{\mbox{d}^4x_1}\int{\mbox{d}^4x_2}L^{1W}(x_1)L^{1W}(x_2)\mid
n,0\rangle{=}\nonumber\\ {=}{-}{Q}^2\langle p,\pi^-\mid{\cal
T}\int\mbox{d}^4x_1\int\mbox{d}^4x_2
\int\frac{\mbox{d}^4P}{(2\pi)^4}\times\nonumber\\
\times{\cal D}_{\mu\nu}(P)\exp[-i
P(x_1-x_2)]{\tilde J}^+_{\mu}(x_1){\tilde J}^-_{\nu}(x_2)\mid
n,0\rangle ,
\label{012} \\
{Q}=\frac{|V_{ud}|\sqrt{G}M_W}{2^{1/4}}. \; \; \; \; \; \nonumber
\end{eqnarray}
Here the $W-$boson propagator
\begin{eqnarray}
{\cal D}_{\mu\nu}(P)=\frac{-ig_{\mu\nu}}{P^2-M_W^2+i\epsilon} ,
\label{dw}
\end{eqnarray}
and the non-linear in $\bm{\varphi}^2$ hadron currents are
defined by eqs. (\ref{+5}), (\ref{pmc}), (\ref{jn})-(\ref{5jp}).
For illustration's sake (only), the amplitude (\ref{012}) could be
depicted by the diagram, \hyperlink{ris2}{fig. 2.}, where the zigzag line figures the
pion interaction with nucleon via $W-$boson. The thin lines around
zigzag are to simulate the infinite number of the virtual pion
propagators (the virtual pions cloud), which appears in calculating
the quantity (\ref{012}), as the lagrangian (\ref{1w}) is
of-principle non-linear in pion field.

The amplitude given by eq. (\ref{012}) could be said to be akin to
the one given by eq. (\ref{1d}), in the sense that they both
describe the processes caused by an intermediate $W-$boson.

The total amplitude of the treated pion emission (the total
$NN\pi^- -$vertex) would be written as the sum of the amplitudes
given by eqs. (\ref{jb}) and (\ref{012}),
\begin{eqnarray}
\langle p,\pi^-\mid S^{tot}_{np\pi}\mid n,0\rangle =
\langle p,\pi^-\mid S_{np\pi}\mid n,0\rangle +
\langle p,\pi^-\mid S^W_{np\pi}\mid n,0\rangle. \label{tot}
\end{eqnarray}
In our approach, the $P-$odd, as well as $P-$conserving, transition
amplitudes are directly dictated by the developed lagrangian
(\ref{4l}), (\ref{ew}). So, there sees no reason to invent a
phenomenological $P-$odd pion-nucleon interaction from general
arguments.

The ensuing evaluations get simplified as all the momenta and
masses involved are rather negligible as compared with the
$W-$boson mass: $p_n ,\, p_p ,\, q_{\pi} ,\, M_{n,p} , \, m \,\ll
M_W$ .

Performing the ordinary transformations we obtain
\begin{eqnarray}
\langle p,\pi^-\mid S_{np\pi}^W\mid n,0\rangle=
i Q^2
\int\mbox{d}^4x_1\int\mbox{d}^4x_2\int\frac{\mbox{d}^4P}{(2\pi)^4}
\frac{\exp[-iP(x_1-x_2)]}{P^2-M_W^2+i\epsilon}\times \qquad \nonumber\\
\times
\Bigl\{ {\bar U}_p\gamma^{\mu}({\mathsf Z}_e+g'\gamma^5)U_n
\exp[ix_1(p_p-p_n)]\times  \qquad \nonumber\\
\times\langle \pi^-\mid\Bigl(1+
\bigl(\pi^-(x_1)\pi^+(x_1)\bm{\pi}^2(x_1)\bigl)\frac{\sin^2z(x_1)}
{\bm{\pi}^2(x_1)}\Bigr)
\Biggl(\frac{\sqrt{2}}{2f}\partial_{\mu}\pi^+(x_2)
\frac{\sin 2y(x_2)}{2y(x_2)} + \qquad \nonumber\\
+ 4f\sqrt{2}\pi^{+}(x_2)
\sum_{a=\pm,0}\partial_{\mu}\pi^a(x_2)
\pi^{-a}(x_2)\frac{1}{y^2(x_2)}\Bigl(1-\frac{\sin
2y(x_2)}{2y(x_2)}\Bigr)\Biggr)\mid 0\rangle +  \qquad \nonumber\\ + {\bar
U}_p\gamma^{\mu}({\mathsf Z}_e
g'\gamma^5+1)U_n\exp[ix_1(p_p-p_n)]\times \qquad \nonumber\\
\times\langle\pi^-\mid 2\pi^0(x_1)
\frac{\sin 2z(x_1)}{2\sqrt{\bm{\pi}
^2(x_1)}}\sqrt{2}\Bigl(\partial_{\mu}\pi^0(x_2)\pi^{+}(x_2)-
\pi^0(x_2)\partial_{\mu}\pi^{+}(x_2)\Bigr)\frac{\sin^2y(x_2)}
{y^2(x_2)}\mid 0\rangle \Bigl\} . \qquad \label{03}
\end{eqnarray}
Here, as well as before, ${\bar U}_p , \, U_n $ stand for the
proton and neutron wave amplitudes, and $M_p , \, M_n $ and $p_p ,
\, p_n $ are their masses
 and momenta. Let us recall the operator functions ${\sin z}{/}{z}
 , \, \cos z , $\, and so on , are treated as power series in
square of the pion field operators
$\bm{\varphi}^2(x)=\bm{\pi}^2(x)$ , and the functions $y(x) ,
\, z(x) $ have been introduced before in Sec. 2 , eqs.
(\ref{il1})-(\ref{2pi}). As the expression (\ref{012}) is a
transition amplitude from pion vacuum $\mid n,0\rangle$ to a state
$\langle p,\pi^-\mid$ containing one negative pion, $\pi^-$ meson,
the integrand in eq. (\ref{03}) incorporates the matrix element
$\langle\pi^-\mid . .
.\mid 0\rangle$ of a product of odd number of the pion field
operators ${\pi}^{a}(x)$ depending on coordinates $x_1$ and $x_2$
of the vertexes $L^{1W}(x_1)$ and $L^{1W}(x_2)$ in eq. (\ref{012}).
According to the Wick-theorem, this matrix element is a sum of
products of the matrix element $\langle \pi^{-}\mid
\pi^{+} (x)\mid 0\rangle =u_{\pi r}\exp[ixq_{\pi}]$ times the
vacuum expectations of the type
\begin{eqnarray}
\langle 0\mid(\partial_{\mu}\bm{\pi}(x_1)\bm{\pi}(x_1))
\bm{\pi}^{2k}(x_1)\bm{\pi}^{2m}(x_2)\mid 0\rangle , \;
\langle 0\mid\bm{\pi}^{2n}(x_1)
(\partial_{\mu}\bm{\pi}(x_2)\bm{\pi}(x_2))
\bm{\pi}^{2k}(x_2)\mid 0\rangle ,  \qquad \label{dd2}
\end{eqnarray}
or the matrix element
 $\langle\pi^-\mid\partial_{\mu}\pi^{+} (x)\mid 0\rangle
=u_{\pi r}(iq_{\mu})\exp[ixq_{\pi}]$ times the vacuum expectations
of the form
\begin{eqnarray}
\langle 0\mid\bm{\pi}^{2n}(x_1)
\bm{\pi}^{2m}(x_2)\mid 0\rangle. \label{dd1}
\end{eqnarray}
 Every one of these vacuum expectations is in turn presented as a
product of two kinds of vacuum expectations: the first one -- a
vacuum expectation of a product of the operators $\bm{\pi}^2(x)$,
associated with one and the same vertex $L^{1W}(x)$ and therefore
depending on one single coordinate $x$, $\langle
0\mid\bm{\pi}^{2n}(x)\mid 0\rangle=F_0$ (the closed thin-line
circles in \hyperlink{ris2}{fig. 2}), and the second one -- a certain vacuum
expectation from those set in eqs. (\ref{dd1}), (\ref{dd2}),
depending on the coordinates $x_1$ and $x_2$ relating to the
vertexes $L^{1W}(x_1)$ and $L^{1W}(x_2)$. (These vacuum
expectations are pictured in \hyperlink{ris2}{fig. 2}  by the bows attached to
zigzag.) It is to keep in mind that $\langle
0\mid(\partial_{\mu}\bm{\pi}(x)\bm{\pi}(x))
\bm{\pi}^{2k}(x)\mid 0\rangle =0$. With using the equations
(\ref{h1})-(\ref{h3}), the vacuum expectations $\langle
0\mid\bm{\pi}^{2n}(x)\mid 0\rangle$ are expressed directly in
terms of $\Delta$ (\ref{d1}) and does not depend on coordinates $x$
at all. Therefore the quantity $F_0$ can be brought out from under
integrating over $x_1$, $x_2$ in eq. (\ref{03}). The vacuum
expectation of a pion field operators product, depending on
coordinates $x_1$ and $x_2$, is expressed in terms of the
quantities
\begin{eqnarray}
\langle 0\mid{\cal T}\pi^{\alpha}(x_1)\pi^{\beta}(x_2)\mid
0\rangle =\delta_{\alpha\beta}\int\frac{\mbox{d}^4k}{(2\pi)^4}
\frac{i\exp[-ik(x_1-x_2)]}{k^2-m^2+i\epsilon} , \label{1z1}\\
\langle 0\mid{\cal T}\partial_{\mu}\pi^{\alpha}(x_1)
\pi^{\beta}(x_2)\mid
0\rangle =\delta_{\alpha\beta}\int\frac{\mbox{d}^4k}{(2\pi)^4}
\frac{k_{\mu}\exp[-ik(x_1-x_2)]}{k^2-m^2+i\epsilon} , \nonumber
\end{eqnarray}
{\it i.e.} through the pion propagators and their derivatives. So it
shows up to be function of the difference $x_1-x_2$. Denoting such
vacuum expectations by $F(x_1-x_2)$, the eq. (\ref{03}) would
generally be represented as a sum of expressions of the form
\begin{eqnarray} {F_0iQ^2}
{\int}\mbox{d}^4x_1{\int}\mbox{d}^4x_2{\int}
\frac{\mbox{d}^4P}{(2\pi)^4}
\frac{\exp[-iP(x_1-x_2)]}{P^2-M_W^2+i\epsilon}\times \nonumber\\
\times F(x_1-x_2)\exp[ix_1(p_p-p_n)+ix_2q_{\pi}] =\nonumber\\
=F_0 i Q^2 \int\frac{\mbox{d}^4k {\tilde F}(k)}{(p_p-p_n-k)^2-M_W^2}
\delta(q_{\pi}+p_p-p_n). \label{2z}
\end{eqnarray}
As $p_n-p_p\ll M_W$, this difference is to be neglected here. The
Furier-transform ${\tilde F}(k)$ of the function $F(x_1-x_2)$, as
constructed by the functions (\ref{1z1}), has a noticeable value at
$k\ll M_W$, as $m\ll M_W$. Therefore eq. (\ref{2z}) can be
transformed to
\begin{eqnarray}
\frac{iF_0 Q^2}{-M^2_W}\int\mbox{d}^4k {\tilde
F}(k)\delta(p_p+q_{\pi}-p_n)=\frac{iF_0 Q^2}{-M_W^2}
\delta(p_p+q_{\pi}-p_n) (2\pi)^4 F(0), \label{3z}
\end{eqnarray}
with accuracy $\sim {m}{/}{M_W}$. We would apparently receive the
same result just putting $x_1=x_2$ in argument of the function $F$
in the expression (\ref{2z}). So, we can from the very first
presume $x_1=x_2$ in calculating the matrix elements
$\langle\pi^-\mid . .
. \mid 0\rangle$ of the products of pion field operators in the
expression (\ref{03}). As understood, due to $ M_W\gg m , \, p_p ,
\, p_n , \, p_{\pi} , $ integrating over the coordinates $x_1 , \,
x_2$ of two vertexes in eqs. (\ref{012}), (\ref{03}) is constricted
into integrating over single variable. (In \hyperlink{ris2}{fig.2}, this would be
reflected by conflating two points $x_1$ and $x_2$.) Then we
eventually obtain from eq. (\ref{03})
\begin{eqnarray}
\langle\pi^-,p\mid S_{pn\pi}^W\mid 0,n\rangle =\frac{|V_{ud}|^2
G}{2f}u_{\pi r}(2\pi)^4\delta(p_p-p_n+q_{\pi})\times\nonumber\\
\times\Bigl\{[g'{\bar U}_p \gamma_{\mu}\gamma^5 q_{\pi}^{\mu} U_n +
{\mathsf Z}_e {\bar U}_p \gamma_{\mu}q^{\mu}_{\pi} U_n]\frac{1}{9}
\langle 0\mid(1+2\cos y )(2+\frac{\sin 2y}{2y})\mid 0\rangle
-\nonumber\\
-[{\mathsf Z}_e g'{\bar U}_p \gamma_{\mu}\gamma^5 q^{\mu}_{\pi} U_n+
{\bar U}_p\gamma_{\mu} q^{\mu}_{\pi} U_n]\frac{1}{3}\langle 0\mid
\frac{\sin^3 y}{y}\mid 0\rangle\Bigr\} . \label{A}
\end{eqnarray}
The vacuum expectations herein are evaluated just alike the
analogous quantities were calculated in the previous sections using
eqs. (\ref{h1})-(\ref{d1}), which results in
\begin{eqnarray}
\langle\pi^-,p\mid S_{pn\pi}^W\mid 0,n\rangle
=\frac{|V_{ud}|^2
G}{2f}(2\pi)^4\delta(p_p-p_n+q_{\pi}) u_{\pi r}{\bar
U}_p\gamma_{\mu}q^{\mu}_{\pi}\times\nonumber\\
\times\Bigl\{[g'\gamma^5+{\mathsf Z}_e]
\frac{1}{9}[2+\exp({-{1}{/}{4P}})\Bigl(\frac{9}{2}-\frac{2}{P}+
\frac{3}{2}\exp({-{2}{/}{P}})\Bigr) + \exp(-{1}{/}{P})]-\nonumber\\
-[1+g'{\mathsf Z}_e\gamma^5]
\frac{1}{4}\exp({-{1}{/}{4P}})(1-\exp(-{2}{/}{P}))\Bigr\}U_n ,
 \label{AP}
\end{eqnarray}
where $P$ is defined by eq. (\ref{p0}). In the physical limit, that
is at $d\rightarrow 4 , \, \Delta\rightarrow
\infty$ in the definition (\ref{d1}) (see \hyperlink{appen}{Appendix A})
, and ${\mathsf Z}_e =3$,
$g'=3g_A$ accordingly eqs. (\ref{1j1}), (\ref{j6}), one obtains
\begin{eqnarray}
\langle\pi^-,p\mid S_{pn\pi}^W\mid 0,n\rangle=
\frac{|V_{ud}|^2G}{2f}(2\pi)^4\delta(p_p-p_n+q_{\pi})\frac{2}{3}
\times \nonumber\\
\times \Bigl\{g_A u_{\pi r}
{\bar U}_p\gamma_{\mu}\gamma^5q^{\mu}_{\pi}U_n + u_{\pi r} {\bar
U}_p\gamma_{\mu}q^{\mu}_{\pi}U_n\Bigr\}. \label{AP0}
\end{eqnarray}

In the lowest $f^2\bm{\pi}^2$-order, when $P^{-1}\rightarrow 0$,
and $g'=g_A$ accordingly eq. (\ref{j5}), ${\mathsf Z}_e =1$ accordingly
eq. (\ref{1j1}), the
 amplitude (\ref{AP}) reduces plainly to the expression (\ref{AP0})
yet without the multiplier ${2}{/}{3}$. This last outcome would
apparently be obtained from the eq. (\ref{012}) if one treated
therein immediately the hadron currents (\ref{5jn}), (\ref{5jp}) in
the lowest $\bm{\varphi}^2$-order:
\begin{eqnarray}
\bfgr{\cal J}_{\varphi \, 5\mu}\approx\frac{1}{2 f}\partial_{\mu}
\bm{\varphi} , \; \; \bfgr{\cal J}_{N \, 5\mu}\approx g_A \bar
N\gamma^{\mu}\gamma^5\frac{\bm{\tau}}{2} N . \label{appr}
\end{eqnarray}
In this case, the amplitude $\langle\pi^-,p\mid S_{pn\pi}^W\mid
0,n\rangle$ would be depicted by the diagram in \hyperlink{ris2}{fig. 2}, yet now
without the thin lines.

 To judge how significant $P$-parity violation is in the
$n\rightarrow p+\pi^-$ process (the $NN\pi-$vertex), we are to
correlate the amplitudes (\ref{AP}), (\ref{AP0}) with the
amplitudes (\ref{jb}), (\ref{jb2}) that describe this transition
without violating $P$-parity.
 The first term in the expression (\ref{AP0}), involving operator
$\gamma^5$, does not violate $P$-invariance, yet causes only a
modification of the parity-conserving pion-nucleon interaction
amplitude, which is rather of no value. The second term, not
including $\gamma^5$ operator, is obviously $P$-non-invariant. The
total transition amplitude (\ref{tot}), with allowance for parity
violating, is then written as
\begin{eqnarray}
\langle p,\pi^-\mid S^{tot}_{np\pi}\mid n,0\rangle =
(2\pi)^4\delta(p_p-p_n+q_{\pi})u_{\pi r}{\bar
U}_p\gamma_{\mu}q^{\mu}_{\pi}\Bigl(\gamma^5
\frac{g_A\sqrt{2}}{2F_{\pi}}+\frac{2}{3}|V_{ud}|^2 G
F_{\pi}\Bigr) U_n .  \qquad \label{tt}
\end{eqnarray}
So, as a matter of fact, we are to compare the quantity
$({\sqrt{2}g_A}{/}{2F_{\pi}})({\bar U}_p
\gamma_{\mu}\gamma^5 q^{\mu}_{\pi}U_n)$, coming from the amplitude
(\ref{jb2}), with one $({2|V_{ud}|^2GF_{\pi}}{/}{3})({\bar
U}_p\gamma_{\mu}q^{\mu}_{\pi}U_n)$ from the amplitude (\ref{AP0}).
Upon utilizing the generally known magnitude of the quantities $g_A
, \, F_{\pi} , \, G$ , we infer that the $P$-parity violating part
constitutes the portion $$ h^W\approx 0.75\cdot 10^{-7}\frac{({\bar
U}_p\gamma_{\mu}q^{\mu}_{\pi}U_n)}{({\bar
U}_p\gamma_{\mu}\gamma^5q^{\mu}_{\pi}U_n)}$$ of the parity
conserving part of the transition amplitude. The total transition
amplitude (\ref{tt}) would directly be obtained if we described the
considered pion emission by the lagrangian (\ref{il1}) with
replacing therein $\gamma^5
\Rightarrow\gamma^5 + 0.75 \cdot 10^{-7} {\mbox{I}} $.

When we were dealing with on-mass-shell nucleons (yet still an
off-mass-shall pion), the amplitude (\ref{tt}) would reduce to
\begin{eqnarray}
\langle p,\pi^-\mid S^{tot}_{np\pi}\mid n,0\rangle = -
(2\pi)^4\delta(p_p-p_n+q_{\pi})\times\nonumber\\
\times u_{\pi r}{\bar U}_p\Bigl(\gamma^5
\sqrt{2}g_{NN\pi}+
\frac{2}{3}|V_{ud}|^2 G F_{\pi}(M_n-M_p)\Bigr) U_n . \label{MM}
\end{eqnarray}
 Then the P-parity violating part would constitute the portion
\begin{eqnarray}
h^W_{(OMS)}\approx0.75 \cdot
10^{-7}\frac{(M_n-M_p)}{(M_n+M_p)}\approx 0.5 \cdot 10^{-10}
\label{hwm}
\end{eqnarray}
of the total transition amplitude. So, in this case, parity
violating is realized to be rather not discernible. With allowance
for eq. (\ref{je}), one would come to this transition amplitude
(the $NN\pi-$vertex) (\ref{MM}) starting with the pion-nucleon
interaction ${\cal L}_{NN\pi}$ (\ref{ja}), replaced
$\gamma^5\Rightarrow\gamma^5 +\mbox{I}\,h^W_{(OMS)}$ therein. It is
to emphasize once again that eq. (\ref{tt}) reduces to eq.
(\ref{MM}) (alike eq. (\ref{jb}) to eq. (\ref{mpn}) does) for the
case of on-mass-shell (free) nucleons, but never for the nucleons
embedded into nuclear matter (in nuclei).

In framework of the approach developed in refs. \cite{DDH,DZ,Hol},
the $P-$odd amplitude of the considered transition $n\rightarrow
p+\pi^-$ is due to the phenomenological interaction
\begin{eqnarray}
{\cal L}^D_{NN\pi}=\frac{h^1_{\pi}}{\sqrt{2}}{\bar N}
[\bm{\tau}\times\bm{\varphi}]_3N , \label{hol}
\end{eqnarray}
added to the interaction (\ref{ja}) (conserving $P-$parity) that
defines the first term in the amplitude (\ref{MM}). There exists a
number of attempts \cite{DDH,DZ,Hol} to evaluate this coupling
$h^1_{\pi}$ using the basic quark-models techniques, but they all
have encountered significant ambiguities. Their results could be
presented in terms of an allowable range for $h^1_{\pi}$,
accompanied by a ``best value" representing the best guess for it.
The interaction (\ref{hol}) causes the transition amplitude
\begin{eqnarray}
\langle p,\pi^-\mid S^D_{np\pi}\mid n,0\rangle =\frac{ih^1_{\pi}}{
\sqrt{2}}\langle p,\pi^-\mid\int{\bar N}[\bm{\tau}\times
\bm{\varphi}]_3N\mbox{d}^4x \mid
n,0\rangle=\nonumber\\
=-h^1_{\pi}\langle p,\pi^-\mid\int{\bar N}(\tau^+\pi^+)N\mbox{d}^4x
\mid n,0\rangle=-(2\pi)^4\delta(p_p+q_{\pi}-p_n) u_{\pi r}{\bar
U_p}U_n h^1_{\pi}. \label{hl1}
\end{eqnarray}
Correlating this quantity with the second term in the amplitude
(\ref{MM}), we obtain
\begin{eqnarray}
h^1_{\pi}=\frac{2}{3}|V_{ud}|^2 G F_{\pi} (M_n-M_p)\approx
10^{-9}\approx g_{\pi}0.026, \; \; (g_{\pi}=3.8\cdot 10^{-8})
\label{hl2}
\end{eqnarray}
which is much smaller than the ``best guess" value asserted in ref.
\cite{DDH}, as well as the others discussed in refs.
\cite{DDH,DZ,Hol}. However it still falls within the ``reasonable
range" estimated in ref. \cite{DDH}. As one infers, inquiring into
the contemporary issues explicated in ref. \cite{Hol}, a reliable
$h^1_{\pi}$ value is not be thought to be definitively obtained
from experimental results by now, in spite of the grate deal of
efforts.

Besides the process just examined (the charge pion emission), we
are to discuss the neutral pion (of-mass-shell) emission : $n \,
[p] \rightarrow n \, [p] + \pi^0$. The corresponding transition
amplitude $\langle n,\pi^0\mid S^W_{nn\pi}\mid n,0\rangle$ is
determined by the interaction of eq. (\ref{1z}) in much the same
way as $\langle p,\pi^-\mid S^W_{pn\pi}\mid n,0\rangle$ was by the
one of eq. (\ref{1w}). It is straightforward to ascertain the
parity violating part of the amplitude $\langle n,\pi^0\mid
S^W_{nn\pi}\mid n,0\rangle$ vanishes, just because the initial and
final nucleons have got the
 equal masses. This is in accordance with the general theorem
\cite{gb} that is valid in the $CP-$conserving limit which we
employ. Apparently, the process involving $\pi^0$ is eliminated
with the phenomenological interaction (\ref{hol}).

\section{Conclusions.}
\label{sec:level6}

As understood, we have successively, step by step come to findings,
relying on the general symmetry principles.

Sufficient trueness of the findings is ensured as we pursue the
consistent way set forth. We describe the low-energy electro-weak
transitions in a pion-nucleon system by the developed effective
lagrangian (\ref{ew}), essentially nonlinear in pion field. All the
parameters emerging in its deriving have been in due course
associated with the observable data.

As based on the adoption of eqs. (\ref{1pi}), (\ref{3u}),
(\ref{1ct}), the low-energy effective lagrangian (\ref{4l}) (as
well as the one (\ref{ew}) ensuing from that) incorporates
 the one- and two-derivative terms only. Each additional derivative
$\partial_{\mu}U$ would enter into the effective pion-nucleon
lagrangian with the coefficient $\sim{1}{/}{F_{\pi}}$ (see, for
instance, \cite{d}). In evaluating the matrix elements of pion and
nucleon interactions, this derivative would become a factor of the
momentum $q$ transferred in interactions. Since we consider the
low-energy processes in which the relation ${q}{/}{F_{\pi}}\ll{1}$
is suggested, the allowance for one- and two-derivative terms in
the effective lagrangian is sufficient with acceptable accuracy.

 By the same reason of treating the low-energy processes, the
fields interactions $L^{em} , \, L^{1W} , \, {\cal
L}_{\varphi\varphi} , \; {\cal L}^{int}_{1N\varphi} , ... $
  themselves are allowed for at the lowest order in the
calculations explicated in Sec. IV , \, V.

 Strong interactions are assured to be adequately allowed for in
the approach presented, which is based on the effective lagrangian
essentially nonlinear in pion field.
 In our treatment, we are dealing with the infinite series in pion
field. It is of crucial value that we have never restricted
ourselves by allowance for a finite number of terms in the power
series in $\bm{\varphi}^2$, yet we summarize all this infinite
series, in treating the considered processes. Just allowance for
all the infinite series in $\bm{\varphi}^2$ provides, as we have
seen, to avoid occurrence of divergence in the developed treatment.
In fine, we dare to assert that the plausible accuracy of our
evaluations (in particular that of the eq. (\ref{hl2}) ) is about a
few tens of per-cents, at least.

All the calculations that have heretofore been discussed are
restricted by allowance for the electro-weak interaction in the
lowest order. Certainly, to analyze the observed data with
 high accuracy ($\sim 1\%$ or better), the radiative corrections
are to be taken into consideration in studying the electro-weak
processes. The approach developed would be of use in a simultaneous
treatment of strong and electro-weak interactions in these
processes.
\hypertarget{appen}
\appendix
\section { }

To deal with the quantities akin to $\Delta$ (\ref{h2}), $\Delta_{\mu\nu}$ (\ref{h0}), an idea is to generalize the dimensionality of the space from four ($4$) to a number $d$ (generally speaking, complex) which is called the regularizing parameter \cite{le75}. The integrands in eqs. (\ref{h0}), (\ref{h2}) are functions of $p^2=\eta^{\mu\nu}p_{\mu}p_{\nu}$, with $\eta^{\mu\nu}$ being a diagonal form with $(\nu-1)$ values $+1$ and one $-1$. Modifying properly the integration measure over momenta, we rewrite $\Delta$ in $d$-dimensions in the Euclidean matric
\begin{eqnarray}
\Delta(d)=\frac{i\mu^{\epsilon}}{(2\pi)^d}\int\frac{\mbox{d}^d p}{p^2+m^2} =\nonumber\\
=\frac{\mu^{\epsilon}}{2^d \pi^{d/2}}\frac{1}{\Gamma(d/2)}\int\limits_0^{\infty}\frac{\mbox{d}(p^2)(p^2)^{d/2-1}}{p^2+m^2}
 \, ,  \; \; \; \epsilon=d-4 , \label{app1}
\end{eqnarray}
where, as usual, $\mu$ is an auxiliary quantity having
dimension of mass, $\Gamma$ is the Euler Gamma-function, and
physical limit corresponds to $d\rightarrow 4$. As seen, an implementation of the dimensional regularization necessitates treating an integral of the type (Euler integral)
\begin{eqnarray}
I(d)=\int\limits_{0}^{\infty}\frac{x^{d/2-1}}{x+1}\mbox{d}x , \label{appii}
\end{eqnarray}
multiplied by $(m^2)^{d/2-1}$. Separating, in the usual fashion, integer part of $d/2$,        \begin{eqnarray}
d/2=\nu_d+n_d , \; {\mathrm{with}} \; n_d=0,\pm 1, \pm 2, ... , {\mathrm{and}} +0\leq \mbox{Re} \, {\nu_d}\leq 1-0, \label{app4}
\end{eqnarray}
the integral (\ref{appii}) is generally expedient to be presented in the form
\begin{eqnarray}
I_{n_d}(\nu_d)=\int\limits_0^{\infty}\frac{x^{\nu_d+n_d-1}}{x+1}\mbox{d}x \equiv I(d). \label{appin}
\end{eqnarray}
Of course, for a given $d$ value, the number $n_d$ is dictated by eq. (\ref{app4}).

The function $I(d)=I_{n_d}(\nu_d)$ (\ref{appii}), (\ref{appin}) is clearly only sensible for
\begin{equation}
+0\leq\mbox{Re} \, {d/2}\leq 1-0, \; {\mathrm{i. e. \; for}} \; n_d = 0 . \label{app3}
\end{equation}

 To deal with the function $I(d)$ (\ref{appii}), (\ref{appin}) at a certain $n_d \neq 0$ corresponding to a given $d$ value, in particular, in a vicinity of the physical value, $d=4$ , we are to resort to the analytic extension (continuation) of the function $I_{n_d=0}(\nu_d)$ to values
 $n_d \neq 0$ (see, for instance, refs. \cite{a1,a2,a3,int}).

For $n_d=0$ the analytic function $I(d)=I_{n_d=0}(\nu_d)$ is calculated as follows. To start with, we write
\begin{eqnarray}
I_{n_d=0}(\nu_d)=\int\limits_0^1\mbox{d}x + \int\limits_1^{\infty}\mbox{d}x = I^{(1)} + I^{(2)}. \label{app0i0}\end{eqnarray}
For $+0\leq x\leq 1-0$, the series expansion is valid
\begin{eqnarray}
\frac{x^{\nu_d-1}}{1+x}=\sum_{k=0}^{\infty}(-1)^k x^{\nu_d+k-1} . \label{appi1}
\end{eqnarray}
As the integral of this sum converges uniformly, both at $x=0$ and at $x=1$, integrating therm-by-therm holds true, which results in
\begin{equation}
I^{(1)} = \sum_{k=0}^{\infty}\frac{(-1)^k}{\nu_d+k} . \label{appi2}
\end{equation}
Upon setting $x=1/z$, the integral $I^{(2)}$ results, in the same way, as
\begin{eqnarray}
I^{(2)}=\sum_{k=1}^{\infty}\frac{(-1)^k}{\nu_d-k}. \label{appi3}
\end{eqnarray}
Recalling the common expansion of the function $\pi\mbox{cosec}(\pi \nu)$ into simple fractions, the expression (\ref{app0i0}) proves eventually to be presented in the form
\begin{eqnarray}
I_{n_d=0}(\nu_d)=\pi(-1)^m\mbox{cosec}[\pi (\nu_d+m)] , \; \; m=0, \pm 1 , \pm 2 , ... \, , \, +0\leq\mbox{Re}\nu_d\leq 1-0 \,  ,
\label{appi0} \end{eqnarray}
which itself is analytic everywhere, except for the values $\mbox{Re}(\nu_d+ m)=0, \pm 1, \pm 2, ... $.
 In writing eq. (\ref{appi0}), the prime relation is used: $\mbox{cosec}\phi = (-1)^n\mbox{cosec} (\phi + n\pi)$.
With plainly implying $m=n_d$ in (\ref{appi0}), this function (\ref{appi0}) itself is obviously understood as analytical continuation (extension) \cite{a1,a2,a3,int} of the analytic function $ I_{n_d=0}(\nu_d) $ to any $n_d\neq 0$ values,  except for the values $\mbox{Re}(\nu_d+ n_d)=0, \pm 1, \pm 2, ... $. Indeed, both functions $I(d)$ and $\pi (-1)^{n_d} \mbox{cosec}[\pi(\nu_d +n_d)]$ are analytic on the strip where they agree, {\it i.e.} at $n_d=0, \; +0\le \mbox{Re}(d/2)\le 1-0$, but the last one is analytic on the wider domain.

 Thus,  upon extending,  the required function
\begin{eqnarray}
  I(d)=I_{n_d}(\nu_d)=\pi(-1)^{n_d}\mbox{cosec}[\pi(\nu_d +n_d)] \label{appi00}
\end{eqnarray}
 proves to be defined on all the $d$-plane, except for the points $\mbox{Re}(d/2)=0,\pm 1, \pm 2, ... $, where the function $\mbox{cosec}$ itself possesses singularities.

It is expedient to clear up a behavior of the function $I(d)=I_{n_d}(\nu_d)$ at $d\rightarrow 0, \pm 1, \pm 2, ...$, especially at $\mbox{Re} d\rightarrow 4$, the dimensionality of the physical space. Let $\mbox{Re}(d/2)\rightarrow 2-0$, that is $d$ tends to the physical value $4$ from below. This corresponds to $\mbox{Re}\nu_d\rightarrow 1-0$ at $n_d=1$ in the analytic extension $I_{n_d}(\nu_d)$ (\ref{appi00}) of the function $I_{n_d=0}(\nu_d)$. Then we arrive at
\begin{eqnarray}
(-1)^1\pi\mbox{cosec}[\pi(\nu_d +1)] \longrightarrow +\infty , \, \mathrm{when} \; \mbox{Re} \nu_d \rightarrow 1-0 .  \label{app1+}
\end{eqnarray}

Yet, if $d$ tends to the physical value $d=4$ from above, $\mbox{Re}(d/2)\rightarrow 2+0$, then in the analytical extension $I_{n_d}(\nu_d)$ (\ref{appi00}) of the function $I_{n_d=0}(\nu_d)$ we have got $\mbox{Re} \nu_d\rightarrow +0$ at $n_d=2$. This case corresponds to
\begin{eqnarray}
(-1)^2\pi\mbox{cosec}[\pi(\nu_d +2)] \longrightarrow +\infty , \, \mathrm{when} \; \mbox{Re} \nu_d \rightarrow +0 .  \label{app2+}
\end{eqnarray}
Eventually, we infer the function $\Delta{(d)}$ (\ref{d1}), being analytic on the $d$-plane, except for the points $\mbox{Re} d/2 =0, \pm 1, \pm 2, ...$, tends to $+\infty$, when $d$ tends to its physical value $d=4$, both from above and from below.

Amenably to the ordinary symmetry consideration, the relation is valid
\begin{eqnarray}
\int\mbox{d}^d p\frac{p_{\alpha}p_{\beta}}{p^2+m^2} = \eta_{\alpha\beta}\frac{1}{d}\int\mbox{d}^dp\frac{p^2}{p^2+m^2} , \label{appsd}
\end{eqnarray}
which immediately yields eq. (\ref{d2}).

\newpage
\begin{center}
.
\vspace{6cm}

\begin{picture}(180,100)(50,0)
\SetScale{2}
\SetColor{Blue}
\SetWidth{1.5}
\LongArrow(140,60)(101.5,60)
\Text(296,126)[]{\Huge${\textBlue\bm{\pi}^+}$}
\Text(150,143)[]{\Huge${\textRed\bm{W}}$}
\SetColor{Red}
\SetWidth{2.0}
\ZigZag(95,60)(40,60){3.5}{6}
\CCirc(95,60){7}{Blue}{Red}
\SetColor{OliveGreen}
\Vertex(40,60){3.0}
\SetWidth{1}
\LongArrow(40,60)(5,80)
\LongArrow(40,60)(5,40)
\Text(0.,85)[]{\Huge${\textOliveGreen\bm{\mu^+}}$}
\Text(0.,165)[]{\Huge${\textOliveGreen\bm{\nu}}$}
\Text(85,95)[]{\Large${\textBlack\bm{x_2}}$}
\Text(195,90)[]{\Large${\textBlack\bm{x_1}}$}
\end{picture}\\
\end{center}

\vspace{-2.2cm}

\begin{center}
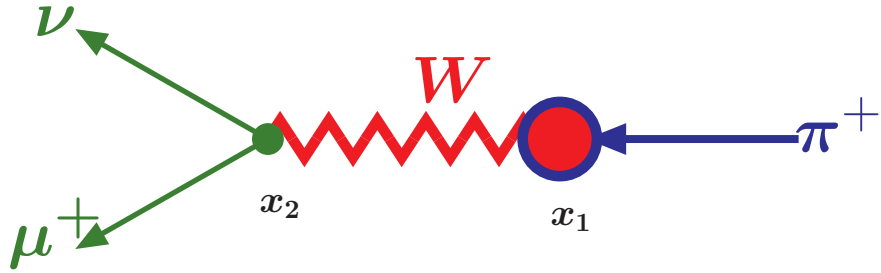
\hypertarget{ris1}{\bf Fig. 1.} Transition amplitude of the decay
$\bm{\pi^+\rightarrow \mu^+ +\nu}$
\end{center}

\begin{center}

\vspace{3cm}

\begin{picture}(360,200)(0,0)
\SetWidth{3.5}
\SetColor{VioletRed}
\ArrowLine(330,60)(160,60)
\ArrowLine(160,60)(20,60)
\SetColor{Red}
\ZigZag(160,60)(160,130){7.5}{5}
\Vertex(160,60){7}
\Vertex(160,130){7}
\SetColor{Blue}
\LongArrow(160,130)(160,190)
\SetWidth{1}
\CArc(160,95)(40,-90,90)
\CArc(160,95)(40,90,270)
\CArc(120,95)(56,-45,45)
\CArc(200,95)(56,135,225)
\CArc(138,95)(50,-48,51.5)
\CArc(182,95)(50,130,225)
\CArc(160,50)(10,0,360)
\CArc(160,45)(15,0,360)
\CArc(160,140)(10,0,360)
\CArc(160,145)(15,0,360)
\Text(300,75)[]{\Huge${\textVioletRed\bm{n}}$}
\Text(50,75)[]{\Huge${\textVioletRed\bm{p}}$}
\Text(185,180)[]{\Huge${\textBlue\bm{\pi^-}}$}
\Text(190,140)[]{\Large${\textBlack\bm{x_2}}$}
\Text(190,45)[]{\Large${\textBlack\bm{x_1}}$}
\end{picture}\\
\end{center}

\vspace{-1.2cm}

\begin{center}
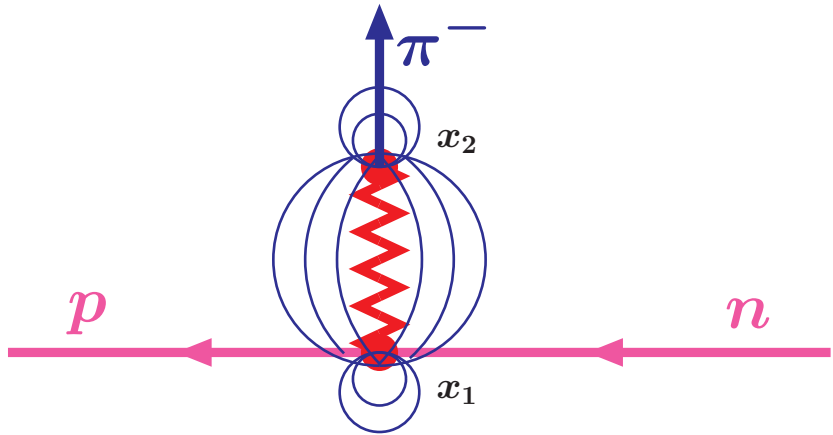
\hypertarget{ris2}{\bf Fig. 2.} The $\bm{n\rightarrow p +\pi^-}$
vertex with P-parity violation.
\end{center}


\vspace{3cm}
{\bf References}
\begin{thebibliography}{99}
\bibitem{d} J.F. Donoghue, E. Golowich and
B.R. Holstein, \href{https://inis.iaea.org/search/search.aspx?orig_q=RN:24065083 }{{\it Dynamics of the Standard Model.} (Cambridge
University Press, Cambridge, UK, 1994.)}
\bibitem{eft} B. R. Holstain, \href{http://www.sciencedirect.com/science/article/pii/S0375947401008284}{Nucl. Phys. {\bf A 689}, 135 (2001).}
\bibitem{cpt}
V.Bernard and U-G. Mei{\ss}ner, \href{http://www.annualreviews.org/doi/full/10.1146/annurev.nucl.56.080805.140449 }{Annual Rev. of Nucl. Part. Science,
{\bf 57}, 33 (2007)};\\
 S.Scheres, \href{http://link.springer.com/chapter/10.1007/0-306-47916-8_2#page-1 }{In: {\it  ``Advance in Nucl. Phys."} {\bf 27}, 201 (2003)
, ed. J.W. Negele and E. Vogt.}
\bibitem{jm} C.N. Yang and R. Mills, \href{http://journals.aps.org/pr/abstract/10.1103/PhysRev.96.191 }{Phys. Rev. {\bf 96},
191 (1954).}
\bibitem{gu} P. Chang and F. G\"ursey, \href{http://journals.aps.org/pr/abstract/10.1103/PhysRev.164.1752}{Phys. Rev. {\bf 164},
1752 (1967)};\\ F. G\"ursey, \href{http://link.springer.com/article/10.1007/BF02860276#page-1}{Nuovo Cim. {\bf 16}, 230 (1960)};\\ F.
G\"ursey, \href{http://www.sciencedirect.com/science/article/pii/0003491661901476 }{Ann. Phys. (N.Y.) {\bf 12}, 91 (1961).}
\bibitem{vp} M.K. Volkov and V.N. Pervushin, \href{http://iopscience.iop.org/article/10.1070/PU1976v019n11ABEH005351/meta;jsessionid=CE1C54823824326BEE5542271FEDCDF8.c1.iopscience.cld.iop.org }{Usp. Phys. Nauk. {\bf
120}, 363 (1976) (in Russia). Soviet Physics Uspekhi, {\bf 19}, iss. 11 (1977).}
\bibitem{kvp} M. K. Volkov and V. N Pervushin, {\it Principle Nonlinear Quantum Theories, Dynamical Symmetries and Meson Physics.} (``Atomizdat", Moscow, 1978.) (in Russia)
\bibitem{dv} D.V. Volkov, \href{http://www.osti.gov/scitech/biblio/4340109}{Part. and Nucl. {\bf 4}, 3 (1973).}
\bibitem{col} S. Coleman, I. Wess and B. Summino, \href{http://journals.aps.org/pr/abstract/10.1103/PhysRev.177.2239 }{Phys. Rev. {\bf
177}, 2239, 2247 (1969).}
\bibitem{sig} M. Gell-Mann and Levy, \href{http://link.springer.com/article/10.1007/BF02859738#page-1}{Nuovo Cim. {\bf 16},
705 (1960)};\\ S. Weinberg, \href{http://journals.aps.org/prl/abstract/10.1103/PhysRevLett.18.188 }{Phys. Rev. Lett. {\bf 18}, 188 (1967).}
\bibitem{eh} G. Ecker and R.W. Honerkamp, \href{http://www.sciencedirect.com/science/article/pii/0550321373902666}{Nucl. Phys. B {\bf 62}, 509 (1973).}
\bibitem{le75} G. Leibbrant, \href{http://journals.aps.org/rmp/abstract/10.1103/RevModPhys.47.849 }{Rev.Mod. Phys. {\bf 47}, 849 (1975)};\\ C. G. Bollini and J. J. Giambiagi, \href{https://www.scopus.com/record/display.uri?eid=2-s2.0-0002672597&origin=inward&txGid=0}{Nuovo Cim. {\bf B 12}, 20 (1972)};\\ C. G. Bollini and J. J. Giambiagi, \href{http://www.sciencedirect.com/science/article/pii/0370269372904832}{Phys. Lett. {\bf 40B}, 566 (1972)};\\
    J. F. Ashmore, \href{http://link.springer.com/article/10.1007/BF02824407}{Nuovo Cim. Lett., {\bf 4}, 289 (1972)};\\ G 't Hoft and W. Veltman, \href{http://www.sciencedirect.com/science/article/pii/0550321372902799}{Nucl. Phys. {\bf B44}, 189 (1972).}
\bibitem{hard} G. H. Hardy, \href{http://www.ams.org/bookstore-getitem/isbn%3D0-8218-2649-2}{{\it Divergent Series}. (Oxford, UK,
1949.)}
\bibitem{ap} S. Scherer and M. R. Schindler, \href{http://link.springer.com/book/10.1007}{{\it Primer for Chiral Perturbation Theory}, Lect. Notes Phys. {\bf 830}. (Springer, Berlin, 2012)}
\bibitem{ll} V.B. Beresteskii, E.M.  Lifshits and L.P.
Pitajevskii, {\it Relativistic Quantum Theory, part I.} (Pergamon,
Oxford, 1971);\\ E.M. Lifshits and L.P. Pitajevskii, {\it
Relativistic Quantum Field Theory, part II.} (Pergamon, Oxford,
1971.)
\bibitem{ad} S.L. Adler and R.F. Dashen, {\it Current
Algebras and Applications to Particle Physics}. (Benjamin,
Inc., New York-Amsterdam, 1968.)
\bibitem{af} V. De Alfaro, S. Fubini, G. Furlan and G. Rosseti, {\it Currents in Hadron Physiks.} (American Elsevier Publishing Company, Inc., New York, 1973.)
\bibitem{pdg} D.E. Groom {\it et all.}, \href{https://inis.iaea.org/search/search.aspx?orig_q=RN:31041263 }{Rew. Part. Phys. (PDG), Eur. Phys. J. C {\bf 15}, 1 (2000).}
\bibitem{bb} E. D. Commins and P. H. Bucksbaum, \href{https://inis.iaea.org/search/search.aspx?orig_q=RN:15054786}{{\it Weak
Interactions of Leptons and Quarks.} (Cambridge University Press,
1983.)}
\bibitem{gtr} M. L. Goldberger and S. B. Treiman, \href{http://journals.aps.org/pr/abstract/10.1103/PhysRev.110.1178}{Phys. Rev. {\bf 110}, 1178 (1958).}
\bibitem{sha} J.-W. Chen and X. Ji, \href{http://www.sciencedirect.com/science/article/pii/S0370269301001009}{Phys. Lett. {\bf B 501}, 209. (2001)};\\ E. Hernandes, J. Nieves and M. Valverde, \href{http://journals.aps.org/prd/abstract/10.1103/PhysRevD.76.033005}{Phys. Rev. {\bf
D76}, 033005 (2007)};\\ P. F. Bedaque, M. J. Sawage, \href{http://journals.aps.org/prc/abstract/10.1103/PhysRevC.62.018501}{Phys. Rev. {\bf
C62}, 018501 (2000)};\\ J.-W. Chen, T. D. Cohen, C. W. Kao, \href{http://journals.aps.org/prc/abstract/10.1103/PhysRevC.64.055206}{Phys.
Rev. {\bf C64}, 055206 (2001).}
\bibitem{DDH} B. Desplanques, J. E. Donoghue, and B. R. Holstein, \href{http://www.sciencedirect.com/science/article/pii/0003491680902171}{
Ann. Phys. (NY) {\bf 124}, 449 (1980).}
\bibitem{DZ} V. M. Dubovik, S. V. Zenkin, \href{http://www.sciencedirect.com/science/article/pii/0003491680902171 }{Ann. Phys. (NY) {\bf 172}, 100 (1986).}
\bibitem{Hol} G. B. Feldman, G. A. Crawford, J. Dubach, and B. R.
Holstein, \href{http://journals.aps.org/prc/abstract/10.1103/PhysRevC.43.863}{Phys. Rev. {\bf C43}, 863 (1991)};\\ N. Kaiser and Ulf-G.
Meissner, \href{http://www.sciencedirect.com/science/article/pii/037594749090359T}{Nucl. Phys. {\bf A510}, 759 (1990)};\\ B. R. Holstein, \href{http://download.springer.com/static/pdf/509/art%253A10.1140%252Fepja%252Fi2009-10798-1.pdf?originUrl=http%3A%2F%2Flink.springer.com%2Farticle%2F10.1140%2Fepja%2Fi2009-10798-1&token2=exp=1449326138~acl=%2Fstatic%2Fpdf%2F509%2Fart%25253A10.1140%25252Fepja%25252Fi2009-10798-1.pdf%3ForiginUrl%3Dhttp%253A%252F%252Flink.springer.com%252Farticle%252F10.1140%252Fepja%252Fi2009-10798-1*~hmac=6167252f6fe4c5b79872dd260ea65fec7628d6a00784a91af8bf474d05add347 }{
EPJ {\bf A 41}, 279 (2009)};\\ B. R. Holstein, \href{http://ac.els-cdn.com/S0375947410004768/1-s2.0-S0375947410004768-main.pdf?_tid=9daf8e60-99bb-11e5-8227-00000aab0f6c&acdnat=1449146858_c4ce89f18cb96617846ee72a79044afc }{Nucl. Phys. {\bf A844}, 160 (2010)};\\ W. C. Haxton and B. R. Holstein, \href{http://www.sciencedirect.com/science/article/pii/S0146641013000288}{Prog. Part.
Nucl Phys. {\bf 71}, 185 (2013).}
\bibitem{gb} G. Barton, \href{http://link.springer.com/article/10.1007/BF02733247#page-1 }{Nuovo Cim. {\bf 19}, 512 (1961).}
\bibitem{a1} E. T. Whittaker and G. N. Watson, {\it A Course of Modern Analysis,} 4th ed. (Cambridge University Press, Cambridge, UK, 1990.)
\bibitem{a2} P. M. Morse and H. Feshbach, {\it Methods of Theoretical Physics, Part 1.} (New York: McGraw-Hill, USA, 1953.)
\bibitem{a3} K. Knopp, {\it Theory of Functions, Part I.} (New York: Dover, USA, 1996.)
\bibitem{int} \href{http://mathworld.wolfram.com/AnalyticContinuation.html} http://mathworld.wolfram.com/AnalyticContinuation.html;\\
\href{http://math.stackexchange.com/questions} http://math.stackexchange.com/questions;\\
S.T.C. Siclos, \href{http://damtp.cam.ac.uk/user/stcs/courses/fcm/sheets/sheet1.pdf} http://damtp.cam.ac.uk/user/stcs/courses/fcm/sheets/sheet1.pdf
\end{thebibliography}
\end{document}